\documentclass[12pt,preprint]{aastex}
\def\ltsima{$\; \buildrel < \over \sim \;$}
\def\simlt{\lower.5ex\hbox{\ltsima}}
\def\gtsima{$\; \buildrel > \over \sim \;$}
\def\simgt{\lower.5ex\hbox{\gtsima}}

\begin{document}

\title{H$\alpha$ Variability in PTFO8-8695 and the Possible Direct Detection 
       of Emission from a 2 Million Year Old Evaporating Hot Jupiter}

\author{Christopher M. Johns--Krull}
\affil{Department of Physics \& Astronomy, Rice University, 6100 Main St.
       MS-108, Houston, TX 77005, USA}
\email{cmj@rice.edu}

\author{Lisa Prato \& Jacob N. McLane\altaffilmark{1,2}}
\affil{Lowell Observatory, 1400 W. Mars Hill Rd., Flagstaff, AZ 86001, USA}
\email{lprato@lowell.edu, jnmclane@astro.as.utexas.edu}

\author{David R. Ciardi\altaffilmark{3} \& Julian C. van Eyken}
\affil{NASA Exoplanet Science Institute (NEXScI), Caltech M/S 100-22, 
       Pasadena, CA 91125, USA}
\email{ciardi@ipac.caltech.edu, vaneyken@ipac.caltech.edu}

\author{Wei Chen\altaffilmark{2}}
\affil{Department of Physics \& Astronomy, Rice University, 6100 Main St.
       MS-108, Houston, TX 77005, USA}
\email{wc2@rice.edu}

\author{John R. Stauffer}
\affil{Spitzer Science Center/Caltech, 1200 East California Boulevard, 
       Pasadena, CA 91125, USA}
\email{stauffer@ipac.caltech.edu}

\author{Charles A. Beichman\altaffilmark{4}}
\affil{Jet Propulsion Laboratory, California Institute of Technology, 4800 
       Oak Grove Drive, Pasadena, CA 91109, USA}
\email{chas@ipac.caltech.edu}

\author{Sarah A. Frazier}
\affil{Department of Physics \& Astronomy, Rice University, 6100 Main St.
       MS-108, Houston, TX 77005, USA}
\email{sarah.a.frazier@rice.edu}

\author{Andrew F. Boden}
\affil{Division of Physics, Math and Astronomy, California Institute of 
       Technology, 1200 E California Blvd., Pasadena, CA 91125, USA}
\email{bode@astro.caltech.edu}

\author{Maria Morales-Calder\'on}
\affil{Centro de Astrobiolog\'ia, INTA-CSIC, ESAC Campus, P.O. Box 78, 
       E-28691 Villanueva de la Canada, Spain}
\email{mariamc@cab.inta-csic.es}

\author{Luisa M. Rebull}
\affil{Spitzer Science Center/Caltech, 1200 East California Boulevard,
       Pasadena, CA 91125, USA}
       \email{rebull@ipac.caltech.edu}

\altaffiltext{1}{Department of Physics \& Astronomy, Northern Arizona 
University, Flagstaff, AZ 86011, USA}
\altaffiltext{2}{Now at Department of Astronomy, The University of
Texas at Austin, Austin, TX 78712, USA}
\altaffiltext{3}{Visiting Astronomer, Kitt Peak National Observatory, 
National Optical Astronomy Observatory, which is operated by the Association 
of Universities for Research in Astronomy (AURA) under cooperative agreement 
with the National Science Foundation.}
\altaffiltext{4}{NASA Exoplanet Science Institute (NExScI), California 
Institute of Technology, 770 S. Wilson Ave, Pasadena, CA 91125, USA}

\begin{abstract} 

We use high time cadence, high spectral resolution
optical observations to detect excess H$\alpha$ emission from the $2 - 3$ 
Myr old weak lined T Tauri star PTFO8-8695.  This excess emission
appears to move in velocity as expected if 
it were produced by the suspected planetary companion to this young star.  
The excess emission is not always present, but when it is, the predicted
velocity motion is often observed.  We have considered the possibility
that the observed excess emission is produced by stellar activity (flares),
accretion from a disk, or a planetary companion; we find the 
planetary companion to be the most likely explanation.  If this is the
case, the strength of the H$\alpha$ line indicates that the emission comes from
an extended volume around the planet, likely fed by mass loss from the
planet which is expected to be overflowing its Roche lobe.  

%Interpreting
%the radial velocity variations of the excess H$\alpha$ emission as coming
%from the planet, we place an upper limit on the mass of the star as 
%$M_* {\rm sin}^3i = 0.535 \pm 0.047$ M$_\odot$, while the planet's
%mass would then be $M_P {\rm sin}^3i = 4.45 \pm 0.34$ M$_{JUP}$.  While
%there is evidence that the orbital inclination of this system varies
%dramatically as a result of nodal precession, the inclination
%at the time of our spectroscopic observations cannot be too low or the 
%stellar mass would be highly inconsistent with its spectral type.  This
%leads to an upper limit for the planet's mass of $\sim 7$ M$_{JUP}$.

\end{abstract}


\section{Introduction} 

The number of known and candidate extrasolar planets continues to grow.
As of 11 May 2016 there are $\sim 3200$ confirmed planets and 
$\sim 2400$
planet candidates\footnotemark
\footnotetext{see http://exoplanets.org; http://planetquest.jpl.nasa.gov;
http://exoplanetarchive.ipac.caltech.edu/}.
The vast majority of these known and candidate planets have been discovered
around middle-aged main sequence stars, and many of these discoveries have
challenged our understanding of planet formation, starting 
with the discovery of the first extrasolar planet orbiting a Sun-like star
\citep[51 Peg b;][]{may95}, the first of the class of ``hot Jupiters." 
The wide variety of extrasolar planetary systems now known has led to 
increased interest and debate over the processes that lead to planet 
formation.  The core nucleated accretion model \citep[e.g.,][]{pol96, bod00,
hub05, lis07} produces Jupiter mass objects slowly as they are 
built up as a result of collisions of dust and ice particles 
which stick together and gradually form larger and larger bodies until
sufficient mass is obtained in order to gravitationally accrete large
amounts of gas from the disk.  The competing gravitational 
instability model \citep[e.g.,][]{kui51, cam78, bos97, bos98, dur07}
posits circumstellar disks which are massive enough to fragment
as a result of their own gravity and form Jupiter mass planets on a
much more rapid timescale than is typical in conventional core accretion
models.  

Both models of planet formation find support and difficulties with 
current observations.  For example, the planet metallicity correlation
\citep{gon97, san04, fis05} is often quoted as evidence in support of 
the core accretion
model.  On the other hand, the direct imaging discovery of massive planets 
on wide orbits \citep[e.g. HR8799b, c, \& d,][]{mar08} has been taken as 
evidence that gravitational instabilities must be important for forming at
least some planets that cannot be easily explained by the core accretion 
scenario \citep{dod09}.  A chief limitation for the core accretion model 
is the timescale involved relative to the lifetime of
circumstellar accretion disks.  Neither the core accretion model nor the
gravitational instability model originally predicted the existence of hot
Jupiters, leading to the need for a mechanism such as migration to move
massive planets from their distant formation sites to final positions in
close to the star \citep[e.g.][]{pap07, lev07}.  As a result, there is a 
significant need to establish the timescale of planet formation and planet 
migration.  The candidate young hot Jupiter studied here could potentially 
set important constraints on our understanding of these processes.

A desirable way to study the planet formation process, its timescale, and the 
role of migration and other phenomena, is to search for planets around young
stars that are in the process of forming their planetary systems.
Detecting this youngest generation ($\sim$ few Myr old) of planets presents
special challenges. The youngest stars, many still surrounded by the
circumstellar material from which planets are presumed to be actively
forming, are mostly located in regions at distances of $>$100~pc.
Thus, these targets are inherently faint and are further obscured and
reddened by material local to the star forming region.
Young pre-main sequence
stars have very strong magnetic fields \citep[e.g.,][]{joh07}
and possess large star spots \citep[e.g.,][]{hat95}.  This makes detection of
extrasolar planets through radial velocity (RV) monitoring difficult because
star spots can introduce periodic RV signals that mimic those produced by 
planetary companions \citep[e.g.,][]{saa97}.  Nevertheless, several radial
velocity searches for planetary mass companions have been or are currently 
being conducted around low mass, relatively young stars, including pre-main 
sequence stars \citep{esp06, pau06, set07, set08, hue08, cro11, cro12, ngu12}.
To date, these studies have yielded one planet around the 100~Myr old 
G1$-$G1.5~V star HD~70573 \citep{set07}.  A planet has also been claimed
around the 10~Myr old classical T Tauri star (CTTSs) TW Hya; however, 
additional study of this object suggests the RV signal from the putative 
planet is actually caused by spot induced radial velocity jitter \citep{huel08}.
Significant spot induced periodic RV variability has been detected in a 
few additional young stars \citep{pra08, mah11}, highlighting the challenges
of this technique when applied to young stars.

Potential planetary mass objects have recently been found around young stars
through direct imaging studies \citep{neu05, luh06, laf08, sch08, ire11, 
kra12, del13, bow13, kra14}.  These objects are typically at orbital 
separations of $\geq 50$ AU with estimated masses of several M$_{JUP}$.
These objects also challenge our models of planet formation, particularly the
core accretion model, as the timescale to form planets at such large distances
in a disk is expected to be about an order of magnitude greater than the 
estimated age of these objects \citep{pol96}.  The mass estimates for these 
objects come from comparing their estimated luminosity and temperature
with theoretical evolutionary models.  The theoretical models are uncertain
at these young ages and the observations required to pin down the luminosity
and temperature have a number of challenges, resulting in considerable
uncertainty in the final mass estimate for a given object.  As an example,
the companion to GQ Lup discovered by \citet{neu05} has mass estimates that
range from 1 M$_{JUP}$ on the low side to $\sim 40$ M$_{JUP}$ on the high
side \citep[e.g.,][]{neu08}.  As we attempt to advance 
our observational and theoretical understanding of planet and brown dwarf
formation, it will be important to obtain strong limits on the mass of
potential companions to young stars.  Such strong mass constraints are
the forte of RV measurements of extrasolar planets, particularly for those
with independent constraints on the orbital inclination.

Transiting extrasolar planets offer several advantages for the study of
sub-stellar mass companions to stars \citep[e.g.][]{cha07}.  Of 
primary advantage is that the
inclination is well characterized allowing for a more certain mass
determination.  Additionally, the radius and hence density of the planet
can be determined, and numerous additional follow up observations are
possible, at least in principle.  Several transit searches for extrasolar
planets around young stars have now been performed \citep{aig07, mil08, 
neu11, vey11, cod13, cod14}.  Very recently, a candidate transiting extrsolar
planet candidate has been reported around a low mass young ($\sim 3$ Myr) star 
(PTFO8-8695) in the Orion OB1a/25-Ori region \citep{vey12}.  This 
discovery paper suggests a planet with a mass $\leq 5.5 \pm 1.4$ M$_{JUP}$ 
and a radius of $1.91 \pm 0.21$ R$_{JUP}$ in a 
0.45 day orbit around a $0.34 - 0.44$ M$_\odot$ M3 \citep{bri05} weak-lined (non
accreting) T Tauri star (WTTS).  The discovery observations noted unusual
changes in the transit light curve from one observing season to the 
next, which \citet{bar13} argue could be the result of mutual 
precession of the stellar rotation axis and the planet's orbital axis
resulting from tidal interaction of the planet with an oblate star.  The
analysis of Barnes et al. suggests a likely planet mass of 3.0 or 3.6
M$_{JUP}$ and radius of 1.64 or 1.68 R$_{JUP}$ depending on the assumed mass
of the star.  Follow-up transit and stellar RV observations by \citet{cia15} 
lend support to this hypothesis.

Here, we report on high spectral resolution optical observations of
PTFO8-8695 densely sampled over a few orbital periods.  We clearly 
detect excess H$\alpha$ {\it emission} that moves in radial velocity 
as predicted by the expected orbit of the companion, providing further 
evidence for the existence the planet.  The H$\alpha$ luminosity associated
with the planet is almost equal to that coming from the star, indicating
that the H$\alpha$ emission volume is substantially larger than the planet
itself.  The most likely explanation is that the planet is losing mass
at a subtantial rate, though at this time we are not able to fully rule
out a small amount of accretion related emission from a very low mass
disk that may remain around this young star.  In \S 2 we describe the
observations of this system, in \S 3 present our analysis of the data,
and in \S 4 provide a discussion of these results, which are summarized
in \S 5.

\section{Observations}

\subsection{HET and Keck Spectroscopy}

Included in the discovery paper of \citet{vey12} is a set of high resolution
echelle spectra of PTFO8-8695 taken at the Hobby--Eberly Telescope 
\citep[HET][]{ram98} and at the Keck I telescope.  At the HET the High 
Resolution Spectrograph \citep[HRS][]{tul98} was employed, while the 
High Resolution Echelle Spectrometer \citep[HIRES][]{vog94} was used at
Keck.  The details of the observations and data reduction procedures can
be found in \citet{vey12}.  The spectral resolution of these observations
is $\sim 15,000$ at the HET and $\sim 60,000$ at Keck.  These data were
used by \citet{vey12} to study the radial velocity variability of 
PTFO8-8695.  Here, we use these observations to investigate the variability
of the H$\alpha$ emission line.

\subsection{McDonald Observatory}

Observations of PTFO8-8695 were taken at the McDonald Obervatory 2.7~meter Harlan 
J. Smith telescope with the Robert G. Tull Coud\'e echelle 
spectrograph \citep{tul95} on UT 15 November 2013.  A 1.2$''$ slit was used
with the E2 grating to give a spectral resolution of R$\sim$61,400 (with
$\sim 2.05$ pixels per resolution element) for
all observations.  Approximately 50 orders with $\sim$100~\AA\ per order
were dispersed across the 2080$\times$2048 Tektronix CCD, covering the
wavelength region $\sim3,400 - 10,900$~\AA.  Integration times for all 
PTFO8-8695 observations at McDonald Observatory 
were 2400~s, and the seeing was $\sim$2$''$ on average.
Because of the faintness of the target \citep[$V = 16.26$,][]{vey12}, the signal
to noise ratio obtained is quite low.  Nevertheless, significant information
can be extracted from the H$\alpha$ emission line of this star.  Table 1
gives a full log of the PTFO8-8695 observations obtained on 15 November 2013.

We also use a spectrum of the dM3e flare star AD Leo as an example of the
H$\alpha$ profile shape of a chromospherically active M star of the same
spectral type with strong emission lines.  
This spectrum was obtained with the same telescope and instrument,
but on UT 8 November 1995.  For this observation, the CCD was placed at
the F1 focus (as opposed to the F3 focus for PTFO8-8695).  A 0.59$''$ slit
was used to observe AD Leo, yielding a spectral resolution $R \sim 120,000$
spread accross $\sim 4$ pixels.  The same CCD was used, resulting in 
only 19 partial ($\sim 23$ \AA) orders, including the one 
containing the H$\alpha$ line, being recorded.

\subsection{Kitt Peak Observatory}

Observations of PTFO8-8695 were also taken at the Kitt Peak National 
Obervatory 4~meter Mayall telescope with the echelle spectrograph on the 
nights of UT $8 - 10$ December 2012.  A 1.5$''$ slit was used with the 
$58.5 - 63^\circ$ grating to give a spectral resolution of R$\sim$25,500
(with $\sim 3.24$ pixels per resoultion element) for all observations.  
The slit length projected to 9.73$''$ on the sky.  Approximately 21 orders 
with $\sim$150~\AA\ per order were dispersed across the 2080$\times$2048 
Tektronix CCD which was binned by a factor of 2 in the cross-dispersion 
direction resulting in 2080$\times$1024 images.  The observed spectra 
covered the wavelength range $\sim5,500 - 8,600$~\AA.  Integration times 
for the Mayall observations of PTFO8-8695 observations ranged from 600 to 
1200 seconds and were typically taken in groups of three exposures with a 
Thorium-Argon lamp exposures taken at the begining of each group.  The seeing
varied during the run but was typically $\sim$2$''$.  Again, 
the signal to noise ratio obtained is relatively low.
A full log of the PTFO8-8695 observations made at Kitt Peak is given in Table 2.

\subsection{Data Reduction}

All spectra were reduced with custom IDL echelle reduction
routines which have been broadly described by Valenti (1994) and Hinkle et 
al. (2000).  The reduction procedure is quite standard and includes bias 
subtraction, flat fielding by a normalized flat spectrum, scattered light 
subtraction, and optimal extraction of the spectrum.  The blaze function of 
the echelle spectrometer is removed to first order by dividing the extracted 
stellar spectra by an extracted spectrum of the flat lamp.  Final continuum 
normalization was accomplished by fitting a low order polynomial to the
blaze corrected spectra in the regions around the lines of interest for
this study.  For the Mayall spectra, there was room on
the CCD where sky spectra are recorded above and below the stellar spectrum.
A sky spectrum was extracted $\sim 3''$ above or below (depending on 
how well centered the star was) the stellar spectrum.  The 
resulting sky spectrum was scaled to match the sky lines away from H$\alpha$ and
was then subtracted from the object spectrum.  As shown below, the features of
interest for this study are much broader than sky lines, so failure to subtract sky
from the McDonald observations should not have any significant impact on the
final results.  The wavelength solution for the McDonald data was determined by 
fitting a two-dimensional polynomial to $n\lambda$ as function of pixel and order 
number, $n$, for approximately 1800 (for the F3 focus) or 100 (for the F1
focus) extracted thorium lines observed from an internal lamp assembly.  The
wavelength solution for the Kitt Peak data was determined for the H$\alpha$
order only and utilized a 3rd order polynomial fit to 13 extracted Thorium
lines in this order.

\section{Analysis}

\subsection{HET and Keck Data}

Figure 1 shows the 4 H$\alpha$ profiles collected at the HET and the 5
H$\alpha$ profiles collected at Keck.  Each profile is labelled by the
relative phase of the suspected planet (phase of zero is mid-transit), and
time runs down in the Figure.  These profiles were collected over 2 months 
in early 2011.  Only one profile was collected on any given night, so at 
least 2 orbital cycles occur between any two of the observed profiles.  To 
aid in keeping track of the time elapsed between these observations, the 
phases include non-zero values for the integer part of the phase which
represents how may orbits have occured since the first of these exposures.
Many of the profiles in Figure 1 show an essentially symmetric emission
profile about line center with a narrow core on top of a broader emission
base, similar to that seen in rapidly rotating, chromospherically active
dMe stars \citep[e.g.][]{jon96}.  Other profiles show significantly red-
or blue-shifted emission in addition to this centered apparent chromospheric
emission.  There is not an obvious relationship between the location of this
excess emission and the predicted velocity position (shown as the red vertical
line) of the candidate planetary companion.  Overall, Figure 1 shows that 
there is substantial H$\alpha$ line profile variability; however, given the
generally large time delay in orbital cycles from one observation to the
next, it is difficult to understand the source of this variability
without the inclusion of datasets with more dense temporal sampling.

\subsection{McDonald Data}

Figure 2 shows all the observed profiles of PTFO8-8695 obtained on 15 November
2013 at McDonald Observatory.  Each profile is identified with the UT time of
the midpoint of the exposure.  Also shown is the velocity position expected 
for the planetary companion based on the ephemeris published in \citet{vey12}.
There is clear H$\alpha$ emission in the 
profile that appears to be moving in velocity space with the expected
position of the planet.  In addition, there is strong, centrally peaked 
H$\alpha$ emission.  
PTFO8-8695 is a WTTS \citep{bri05, her07} and as such its H$\alpha$ emission
is expected to be chromospheric in origin, that is, it is believed to be
produced by the magnetic activity of the star itself \citep{ber89}.  In
order to estimate the stellar contribution to the observed line profiles
we used two different observed profiles from Figure 2.  In the profile taken
at UT 5:43, the planetary emission appears to be confined to the red side
of the line profile.  The profile taken at UT 10:32 is the one in which the
planet is expected to be the most blue-shifted, and the excess emission 
appears to be confined to the blue side of the line profile.  The red side
of the line profile in both the UT 10:32 and UT 11:17 are nearly identical, 
further indicating the potential planetary emission is confined to the blue
side of the profile.  Therefore, to estimate the stellar chromospheric 
component of the line, we take the blue side of the UT 5:43 profile and 
combine it with the red side of the UT 10:32 profile to get the final profile
shown in Figure 3.  We measured the H$\alpha$ equivalent width of this
stellar emission, finding a value of $10.49 \pm 0.21$ \AA.  Also shown in
Figure 3 is the line profile of AD Leo, a dM3e flare star rotationally
broadened to the same $v$sin$i$ \citep[80.6 km s$^{-1}$][]{vey12} observed
in PTFO8-8695.  \citet{bar13} predict that the apparent $v$sin$i$ of 
PTFO8-8695 will change by $\sim 13$\% as the result of precession of the 
stellar rotation axis.  This effects was looked for by Yu et al. 
(2015) and was not seen, though they only had two observing epochs.  The
exact value we use for $v$sin$i$ may be slightly off; however, we see no
evidence that this is the case.
The width of the rotationally broadened AD Leo 
spectrum is similar to the reconstructed ``chromospheric" profile of 
PTFO8-8695, but
is weaker.  Multipling the AD Leo emission component by a factor of 2.4
leads to the smooth solid profile in Figure 3 which provides a reasonably
good match to the PTFO8-8695 profile, suggesting that the profile presented
in Figure 3 is a good representation of the stellar component of the 
line profile.

Figure 4 shows each of the observed profiles from Figure 2 with the stellar
profile from Figure 3 (the histogram) subtracted off.  The leftover emission
from this
subtraction process could be entirely due to the planetary companion.  
Examining Figure 4 shows that the centroid of this emission tracks very well
the predicted velocity position of the planet.
However, there is a slight hint that the H$\alpha$ emission
is not quite at the predicted velocity from the ephemeris.  In order to 
characterize this apparent planetary emission, we measure the H$\alpha$
equivalent width of each profile in Figure 4, as well as the velocity centroid
of the emission and the velocity width of each profile.  All these values are
reported in Table 1.  For the velocity centroid we compute the flux
(above the continuum) weighted
mean velocity from the data.  For the line width, $\sigma_D$, we use
a measure of dispersion given by
$$\sigma_D = \Bigl( { \Sigma (v - v_o)^2 (F_\lambda - 1) \over \Sigma (F_\lambda - 1) } \Bigr)^{1/2}$$
where $v$ is the velocity of each channel in the continuum normalized spectrum
difference spectrum $F_\lambda$ from Figure 4, and $v_o$ is the previously 
determined velocity centroid.  This measure of 
line width has the advantage of being purely emipirical and does not rely on
fitting any particular functional form to the data.  In the case of a 
Gaussian profile, this $\sigma_D$ is equal to $\sigma$ in the standard 
Gaussian formula.  There is some indication that the dispersions of the
measured profiles are somewhat larger for the times when the planetary 
velocity position is changing the most rapidly.  This may indicate that
there is some smearing of the profile from the relatively long exposure
times compared to the expected orbital period.

\subsection{Kitt Peak Data}

The first two nights (8-9 December 2012) of the Kitt Peak observations of 
PTFO8-8695 are shown in Figures 5 and 6.  In the figures, the individual 
profiles are shown in the black histograms, and the average nightly profile is 
overplotted in red on each of the individual profiles.  In contrast to the
McDonald observations shown in Figure 2, there is no clear excess emission
observed at the predicted velocity position of the planetary companion to 
the star.  Some weak variation in the H$\alpha$ line is observed on both 
nights; however, the velocity positions of these changes appear well within
the range expected for the stellar chromospheric emission.  These variations
are likely caused by spatially localized chromospheric emission features on
the surface of this magetically active WTTS, and are probably unrelated to
the potential planetary companion.  The H$\alpha$ emission on these two
nights, presumably stellar in origin, is somewhat weaker than
seen in the McDonald data: the emission equivalent widths are $8.33 \pm 0.03$
\AA\ and $7.93 \pm 0.06$ \AA\ on 8 and 9 December 2012 respectively.  Since
no excess H$\alpha$ emission is clearly detected on these nights, we do
not attempt to measure any quantities related to this and so no data is
reported for the excess equivalent width, velocity centroid, or line width
in Table 2.

On the third night of the Kitt Peak run, the behavior of the H$\alpha$ line
in PTFO8-8695 became much more active, showing similar behavior to that 
seen in the McDonald data.  The H$\alpha$ profiles from this night
(10 December 2012) are shown in Figure 7, where starting with the fourth
exposure (UT 5:53) obvious excess emission was detected in the H$\alpha$ 
line that again appears to move with the velocity of the planetary companion 
predicted by \citet{vey12}.  
We estimate the stellar chromospheric emission using the profile obtained
at UT 7:34 on this night when the planet is predicted to be most red-shifted.
To minimize the effects of potential stellar H$\alpha$ fluctuations such
as those seen in Figures 5 and 6, we use this profile to estimate the
stellar component of the line since it is close in time to the profiles 
showing the excess emission.  The UT 7:34 profile is reflected about 
zero velocity to create an estimate of the stellar chromospheric profile.
This estimated chromospheric profile has an equivalent width of 
$8.57 \pm 0.12$ \AA.

In an effort to isolate the non-stellar excess emission, we subtract the
estimated chromospheric from all the spectra observed on night 3, and 
the resulting profiles are shown in Figure 8.
We then performed the same
measurements of excess emission equivalent width, velocity centroid, and
line width as described above for the McDonald data.  These values are
given in Table 2.  For the first 2 observations on this night,
we do not report the excess emission velocity centroid or line
width.  Both of these quantities are computed by effectively dividing by
the excess equivalent width.  Because the excess emission is very weak and
not securely detected in these first two observations, the values and
uncertainties in the velocity centroid and line width become extremely
large and provide no constraints on any of the analysis.

\subsection{Revisiting the HET and Keck Data}

As mentioned in \S 3.1, the HET and Keck data collected in 2011 do not
show any obvious relation to the predicted velocity position of the planet.
This can be made more clear by attempting a stellar subtraction similar
to that done to produce Figures 4 and 8.  For these figures, we were able
to use profiles from the same night to represent the stellar component,
but in the case of the HET and Keck data there is only one observation
per night and there is substantial variation from one observation to the
next.  Since the profiles on the first two Kitt Peak nights show no
apparent excess emission, we take the average of these 21 spectra to 
represent the stellar emission.  This stellar component appears to vary in
strength between the McDonald and Kitt Peak observations, and examinations
of the profiles in Figure 1 suggest it does here as well.  Therefore, we 
scaled this stellar component so that it matched the observed profiles in
Figure 1 as best it could, but such that the stellar profile was never above
the observed profile.  The implicit assumption is that the observed profile
is composed of a stellar (including a chromosphere) component plus potentially
an excess emission component on top of this.  Once we subtract the scaled
stellar profiles, only the excess emission is observed and these profiles
are shown in Figure 9.  While some profiles appear to subtract nearly to
zero (e.g. phase 127.792), there is substantial excess emission in many of 
them.  As mentioned above, in some cases the excess emission component is
close in velocity position to the predicted velocity of the planet, while
in other cases it is far away and shows no obvious connection to the suspected
planet.  However, given that substantial time elapsed between each of 
the spectra shown in Figure 9 (and 1), we can not be certain how these 
profiles evolved.  Below, in \S 4.3.4, we offer an interpretation that 
attempts to account for all the observed profile shapes from this star.

\subsection{Orbital Fits}

The observations of the radial velocity variations of the excess H$\alpha$ 
emission recorded at McDonald Observatory and at Kitt Peak, densely sampled
in time, lend themselves to an exploration of possible orbital motion.
Figure 10 shows the measured velocity centroids of the excess H$\alpha$ 
emission against orbital phase based on the ephemeris in \citet{vey12}.
While \citet{bar13} revise the ephemeris of PTFO8-8695, they do not provide
a unique solution; the results depend on the assumed stellar mass.  However,
the ephemerides of Barnes et al. result in a maximum phase offset of 0.026
for the data presented here, which is likely negligible compared to other
uncertainties in the measurements and analysis discussed below.  Therefore,
we use the ephemeris of \citet{vey12} throughout and plot (dashed line)
the predicted velocity curve of the planet in Figure 10 
using their orbital parameters, which assume a circular
orbit.  The value of $\chi^2$ for this velocity curve is 607, and the
reduce $\chi^2$ value is $\chi^2_r = 41$.
The measured velocity variations generally track the expected 
orbital motion of the planet, with the McDonald observations somewhat
closer to the predicted curve than those from Kitt Peak.  We discuss below
the possibility that the excess H$\alpha$ emission includes components 
not purely in orbit with the planet which might lead to this difference; 
however, for completeness, we proceed here assuming these velocity
variations are the result of orbital motion.

The radial velocity observations presented in Figure 10 were obtained in
2012 and 2013, while the ephemeris in \citet{vey12} is based on transit
data taken in 2009 and 2010, so there is some uncertainty in the predicted
phasing of the radial velocity observations presented here.
\citet{cia15} analyze additional transit data for PTFO8-8695 including
photometry from the {\it Spitzer} satellite.  The transits recovered 
in \citet{cia15} were offset from the predicted times using the discovery
ephemeris, but were all within the original ephemeris uncertainty from
\cite{vey12}.  We can use the offset to the observed {\it Spitzer} transit
(data obtained in April 2012) as an estimate of the uncertainty in the 
ephemeris when comparing the predicted phases of the radial velocity 
observations in Figure 10.  The dash-dot curve in this plot shows the
predicted radial velocity curve of the planet using the orbit determined
in \citet{vey12}, shifted to match the transit midpoint observed by
{\it Spitzer} in 2012 \citep{cia15}.  The shift is only 0.026 in phase.
This emphemeris fits the H$\alpha$ RV variations more poorly, giving
$\chi^2 = 889$ and $\chi^2_r = 60$.  This may indicate a problem with the
{\it Spitzer} transit time determination or, as discussed in \S 4, may be
be due to the H$\alpha$ emitting gas not strictly moving with the candidate
planet around this star.

Assuming that the excess H$\alpha$ RV variations shown 
in Figure 10 result from the orbital motion of the planet around PTFO8-8695, we
fit a sine wave (circular orbit) to the RV points (solid 
line in the figure; $\chi^2 = 570$ and $\chi^2_r = 41$), 
obtaining a velocity semi-amplitude of $211.9 \pm 22.1$ km s$^{-1}$.
Clearly, the pure sine wave fit and predicted sine curves (dashed and dash-dot
lines) of Figure 10 are not
good fits to all the observed RV variations of the excess H$\alpha$ emission, 
though they do a fairly good job of fitting the data from McDonald Observatory
in 2013.  We can
obtain a better fit using a full Keplerian curve, allowing the orbit to
be eccentric.  This fit is shown in the dotted line of Figure 10 
($\chi^2 = 177$ and $\chi^2_r = 16$).  In addition
to fitting the RV points, we used the transit time as an additionl constraint.
To do so, we assume the transit midpoint of \citet{vey12} as the observed
transit time and assign an uncertainty equal to the phase offset between
this original transit time and that determined from the {\it Spitzer}
observations of \citet{cia15}.  The resulting velocity semi-amplitude
is $K {\rm sin}i = 268.1 \pm 13.4$ km s$^{-1}$ with an eccentricity of
$e = 0.35 \pm 0.12$.  While such a large eccentricity brings the putative
planet closer to the star, the planet is still outside the star at 
periastron for the largest stellar radius (1.07 R$_\odot$) found by either
\citet{vey12} or \citet{bar13}, though nominally the surface of the planet
would pass within one planetary radius of the stellar surface.  We suggest 
below that there may be significant 
non-orbital contribution to the excess H$\alpha$ RV variations, so a full 
orbital analysis is likely not warranted.  As a result we do not report other
parameters of the orbital fit.  We note that using equation (18) of 
\citet{gu03} that the planet's orbit should be circularized in $10^3 - 10^5$
yrs depending on the exact value of the planetary quality factor $Q^\prime_p$,
so we do not expect an eccentric orbit for the suspected planet, unless that
eccentricity is being excited by a third body in the system.  Below we
argue that the excess H$\alpha$ emission observed in the Kitt Peak spectra
may be particularly affected by non-orbital motion, so we perform a full
Keplerian fit to the McDonald RVs only, again using the transit time as
a constraint.  This fit is shown in the dash-triple dot line in Figure 10 and
gives $K {\rm sin}i = 196.2 \pm 5.6$ km s$^{-1}$ with an eccentricity of
$e = 0.02 \pm 0.05$.  The total $\chi^2 = 14$ which is substantially
improved, while $\chi^2_r = 7.0$ which is also an improvement, but only
modestly so due to the small number of degrees of freedom given only 6
RV data points to which the orbit is fit.

\section{Discussion}

We have detected variable H$\alpha$ emission from the young transiting planet
candidate PTFO8-8695.  This object sometimes shows a component of H$\alpha$
emission that appears to be in excess to the stellar chromospheric emission.
At times, the excess H$\alpha$ appears at a random phase relative
to the expected velocity position of the claimed planetary companion (Figure 1
and 9).
However, at other times the excess H$\alpha$ emission appears to move 
in wavelength as would be expected if it were produced by the suspected
planetary companion.  This raises the intriguing possibility that the excess
H$\alpha$ emission is associated with the planetary companion.
In general though, there are at least three potential
sources of this variable H$\alpha$ emission that should be considered.  The
emission could be associated with 1.) the star itself, 2.) accretion flows
from a tenuous disk, or 3.) it could be directly related to 
a low mass companion. 

\subsection{A Stellar Source: Magnetic Activity and Flaring}

PTFO8-8695 is a young T Tauri star, and as such is expected to be very
magnetically active.  Such stars produce variable chromospheric H$\alpha$ 
emission \citep[e.g.,][]{hat95}.  If all of the observed H$\alpha$ emission 
is taken to be from the star, then the measured emission equivalent width 
when it is active would be $\sim 18$ \AA .  This would give PTFO8-8695 a 
log($L_{H\alpha}/L_{bol}$) = -2.92 which would make it stronger than every 
other magnetically active M star in the sample of \citet{haw96}.  Additionally,
the rotationally broadened and enhanced chromospheric profile of the dMe
flare star AD Leo matches well the core of the H$\alpha$ line from 
PTFO8-8695 (Figure 3).  While there is clearly some variability in this
core as seen in Figures 5 and 6, the strong, highly Doppler shifted excess
H$\alpha$ emission seen in Figures 2 and 7 cannot be explained by 
chromospheric emission on the surface of the star.  

However, there is a
potential that a stellar flare could produce such Doppler shifted 
emission, at least in principle.  Stellar flares in dMe stars often produce
significant, nearly symmetric line broadening at the base of H$\alpha$.
The resulting line profile routinely shows the standard narrow chromospheric
emission on top of a very broad (FWHM of a few hundred km s$^{-1}$) base
of emission \citep[e.g.][]{eas92, jon96} reminisent of some of the weaker
line profiles shown in Figure 1 (e.g. at phases 4.918, 7.127, 22.699, \& 
127.174).  It is very rare in a flare to see dramatically asymmetric emission 
with a highly red- or blue-shifted component nearly equal in strength to the 
central chromospheric emission as seen in the profiles of Figures 2 and 7.

T Tauri stars in general are known to
flare \citep[e.g.,][]{gah90, gue99}.  A potentially better analog of the
type of variable H$\alpha$ emission expected from chromospheric emission and
flaring on PTFO8-8695 is the WTTS V410 Tau, with a $v$sin$i$ = 77.7 km s$^{-1}$ 
\citep[e.g.,][]{car12} compared to the measured $v$sin$i$ = $80.6 \pm 8.1$ 
km s$^{-1}$ for PTFO8-8695 \citep{vey12}.
%, though this value may fluctuate by
%$\sim 13$\% \citep{bar13}.  
The H$\alpha$ emission equivalent width 
($< 3$ \AA\ with a typical value $\sim 1-2$ \AA ) on V410 Tau
\citep[e.g.,][]{hat95, fer04, mek05} is weaker than seen on PTFO8-8695,
though V410 Tau has an earlier spectral type which raises the continuum
level without necessarily affecting the strength of the chromospheric
emission.  V410 Tau has been observed to flare in a number of studies.
Outside of flares, the H$\alpha$ line of V410 Tau is fairly symmetric,
relatively narrow, and is similar in shape to the chromospheric H$\alpha$
profiles for PTFO8-8695 seen in Figures 3, 5, and 6 \citep{hat95, fer04, ske10}.
The H$\alpha$ line of V410 Tau can grow much stronger and broader during
a flare, and also show asymmetries; however, the observed asymmetries 
seen during flares do not show excess emission with apparent
peaks shifted out to greater than $\pm 200$ km s$^{-1}$ \citep{hat95, ric11}
as seen here in PTFO8-8695.  The typical pattern in a flare is for the
line to very rapidly (times scale of a few minutes) strengthen and broaden
with only a slight asymmetry developing.  The strength and width of the line
then decay exponentially with a time scale of $\sim 1$ hour for strong 
flares \citep[e.g.,][]{fer04}.  This is not the temporal behaviour observed
in PTF08-8695.  There is at least one additional piece of evidence against 
the flaring interpretation for the excess H$\alpha$ emission seen in 
PTFO8-8695.  Whenever V410 Tau shows flare emission in H$\alpha$, significant
\ion{He}{1} 5876 \AA\ emission also appears.  This \ion{He}{1} line is covered
in the echelle formats of both our McDonald and Kitt Peak data.
We have searched 
both datasets for evidence of this emission, including co-adding the 
spectra when the H$\alpha$ emission appears stationary (UT 9:44 to 11:17
for McDonald; UT 6:45 to 8:25 for Kitt Peak) to increase the signal to
noise.  No evidence of \ion{He}{1} emission is seen.  Lastly, if the observed
excess H$\alpha$ emission seen in Figures 2 and 7 were the result of a
stellar flare, it would be a remarkable coincidence that the flare induced
asymmetry just happened to appear at and move with the same velocity position
in the
line profile as that expected for the planetary companion.  In particular, the
motion shown in Figures 2 and 4 where the excess emission first appears
strongly on one side of the line profile and then moves to the other side
has not to our knowledge been observed in the H$\alpha$ emission of flare
stars.  Flares have been
observed on PTFO8-8695 \citep{vey12, cia15}, and while flares on this star
likely will produce changes in the strength and shape of the H$\alpha$
emission line, we believe all the points described above argue strongly 
against a purely stellar origin.

\subsection{A Disk Accretion Source}

Another possibility is that the H$\alpha$ emission arises from material 
accreting onto the {\bf star} from a tenuous disk that may still surround
PTFO8-8695.  While this star is classified as a WTTS, it is at the boundary 
between WTTSs and accreting CTTSs \citep{bri05}, although no dust is
evident in its infrared spectral energy distribution, including {\it
Spitzer} data out to 24 $\mu$m \citep{her07}.  If we take the average excess 
H$\alpha$ equivalent width and estimate an accretion rate using the 
empirical calibrations in \citet{fan09}, we find 
a value of $\sim 3 \times 10^{-10}$
M$_\odot$ yr$^{-1}$ which is relatively low compared to the full sample in
Fang et al.  If there is weak accretion from a tenuous disk, 
it is probable that such a disk would be detected in infrared emission as
even very tenuous disks which feed very low accretion rates on the order
of 10$^{-11}$ M$_\odot$ yr$^{-1}$ produce a detectable near-IR excess
\citep{gil14}.  This accretion rate is significantly less than the value 
estimated above, suggesting that if the H$\alpha$ emission was resulted from
disk accretion, PTFO8-8695 should show an IR excess unless there has been 
substantial grain growth around this 2-3 Myr old star which has removed 
essentially all the small grains.  

If PTFO8-8695 is accreting material onto the star
from a gas disk devoid of small grains, the excess
H$\alpha$ emission may result entirely from the accreting material whether
or not there is a planetary companion present.  This accretion related 
emission would presumably be similar to H$\alpha$ emission seen in 
other CTTSs, many of which also have close companions.  If there is a low mass
companion to PTFO8-8695, accretion from a disk may be through
accretion streams such as those proposed by \citet{art96} 
\citep[see also][]{gun02}.  At this time, it is not known if a 
planetary mass companion can excite accretion streams such as those modeled
by \citet{art96} and \citet{gun02}.  A few CTTSs binaries are thought to 
potentially be accreting through accretion streams.  These include DQ Tau 
\citep{mat97, bas97}, UZ Tau E \citep{jen07}, AK Sco \citep{ale03}, 
KH 15D \citep{ham12}, and the eclipsing binary system CoRoT 223992193 in 
NGC 2264 \citep{gil14}.  None of these stars shows the type of H$\alpha$ 
variations seen in PTFO8-8695 where the accretion related emission appears 
to move from one side of the line profile to the other as it spirals onto 
one or both of the stars.  This type of line profile behavior is also not
seen in the H$\alpha$ profile variations
of single CTTSs in extensive studies of their line profile variability 
\citep[e.g.,][]{gia93, joh95a, joh95b, joh97, oli98, ale01}, nor is it
predicted from theoretical models of magnetospheric accretion such as those
shown in \citet{kur13}.  
%Finally, similar
%to the argument given above, if the observed excess H$\alpha$ emission seen 
%in Figures 2 and 7 is indeed due to weak accretion from a disk, it is a 
%remarkable coincidence that this emission just happens to appear at the 
%same velocity position in the line profile as that expected for the 
%planetary companion.  
While we cannot completely rule out accretion from 
a tenuous disk as the source of the excess H$\alpha$ emission observed in 
PTFO8-8695, we argue that this is not the most likely explanation of the
observed emission.  Deep mid IR or millimeter continuum observations,
or a deep search for close circumstellar disk gas emission 
(e.g. H$_2$ emission, see
France et al. 2012), could shed light on whether there is a tenuous disk
around this star feeding accretion onto it.

\subsection{A Planetary Companion Source}

Particularly given the radial velocity variations of the excess emission
component of the H$\alpha$ line, the most likely explanation is that 
this emission arises from an orbiting companion.  

{\bf \subsubsection{Mass Estimates}}

Assuming the companion 
hypothesis, we can use these measured RV variations with the orbital fits 
performed above to estimate the mass of the system.  \citet{vey12} did not 
positively detect RV motion of PTFO8-8695 
itself; however, if we adopt their upper 
limit on the reflex motion of the star ($2.13 \pm 0.12$ km s$^{-1}$), we can
use Kepler's laws to find an upper limit for the total mass in the system of 
$(M_* + M_P) {\rm sin}^3i = 0.456 \pm 0.082$ M$_\odot$ assuming a circular
orbit (solid curve in Figure 10).  
Alternatively, if we use the full Keplerian solution and
again apply the upper limit on the stellar reflex motion of the star
from van Eyken et al. (assuming that this is the upper limit for the velocity
semi-amplitude, $K$), we obtain an upper mass limit for the system of
$(M_* + M_P) {\rm sin}^3i = 0.539 \pm 0.047$ M$_\odot$.  Finally, if we
restrict ourselves to the fit to the McDonald only data, we find an upper
limit to the mass of $(M_* + M_P) {\rm sin}^3i = 0.362 \pm 0.018$ M$_\odot$.
We can then use
these total mass estimates with the observed ratio of the companion and
stellar RV amplitudes, again adopting the upper limit on the stellar
RV amplitude of the star from van Eyken et al., to estimate the mass of the
companion.  Using the parameters from the circular orbit fit
gives $M_P {\rm sin}^3i = 4.75 \pm 0.56$ M$_{JUP}$, while using the 
parameters from the full Keplerian fit gives $M_P {\rm sin}^3i = 4.45 \pm 0.34$
M$_{JUP}$, and the McDonald only fit giving 
$M_P {\rm sin}^3i = 4.07 \pm 0.45$ M$_{JUP}$, all
clearly in the planetary range.  
\citet{vey12} determine an orbital inclination of 
$61.^\circ8 \pm 3.^\circ7$ which then gives a total
mass for the system of $0.666 \pm 0.108$ M$_\odot$ for the circular orbit,
$0.787 \pm 0.095$ M$_\odot$ for the eccentric orbit, and 
$0.529 \pm 0.027$ M$_\odot$ for the McDonald only orbital fit.
The planet mass then becomes $6.94 \pm 0.92$ M$_{JUP}$ for the circular
fit, $6.50 \pm 0.76$ M$_{JUP}$ for the eccentric fit, and 
$5.95 \pm 0.66$ M$_{JUP}$ for the McDonald only fit.
Again, this is an upper limit to the planet's mass given that 
the RV variations measured for the star are only upper limits on the reflex 
motion of the orbit.  Indeed, \citet{cia15} present new K band based RV
measurements of the star, estimating a new stellar RV semi-amplitude of
$K = 0.370 \pm 0.333$ km s$^{-1}$, which would lower the estimate
of the planet's true mass to $\sim 1.1$ M$_{JUP}$.  The true stellar RV 
variations could be even smaller which would imply a still lower
planet mass, though if the planet's mass is too low its Roche
radius would become smaller than the planet and it should then be losing
substantial mass.

The estimates given above for the actual mass of the planet depend on 
the inclination of the planet's orbit, which is uncertain.
\citet{vey12} determine an orbital inclination of $61.^\circ8 \pm 3.^\circ7$;
however, \citet{bar13} argue that this inclination changes over time as a
result of nodal precession of the orbit.  They find the value of the orbital
inclination can change from $\sim 25^\circ$ to $90^\circ$ with a period
in the range of 300-500 days.  Thus, there is the possibility the planet's
orbital inclination at the time of the observations presented here was
significantly different than the $61.^\circ8$ assumed.  Assuming 
$i = 61.^\circ8 \pm 3.^\circ7$, the stellar mass inferred from the
circular orbit fit is $M_* = 0.659 \pm 0.108$ M$_\odot$.  The effective 
temperature of PTFO8-8695 is 3470 K \citep{bri05}.  Using the pre-main
sequence evolutionary tracks of \citet{sie00}, this corresponds to a
mass of $\sim 0.35$ M$_\odot$ at an age of 2 Myr [the tracks of Baraffe et al.
(2015) give a mass of $\sim 0.32$ M$_\odot$ at the same temperature and age]. 
Thus, our measured mass is larger than the predicted masses by 
$\sim 3\sigma$.  It is well established that theoretical pre-main sequence 
tracks underestimate the true mass of young stars at these masses 
\citep{hil04}.  According to these authors, pre-main sequence tracks routinely
predict a mass that is 30\% lower than the true mass.  Thus, if the true mass
is 0.659 M$_\odot$, we might expect evolutionary tracks to give a mass of 
0.461 M$_\odot$.  With an uncertainty of 0.108 M$_\odot$, the actual predicted
masses from the evolutionary tracks are 1.0 -- 1.3 $\sigma$ lower than this
value.  Thus, while our inferred mass is higher than the predicted mass, 
the difference is within the expectations given the known systematic
trends in the evolutionary tracks at these low masses.  No matter what the
true inclination is, our measured mass will be larger than the mass inferred
from evolutionary tracks.  For example, if the inclination is 90$^\circ$, the
measured mass would be 0.451 M$_\odot$.  While this is larger than the mass
inferred from the tracks, it is larger by $\sim 30$\% which is the typical
difference found for young stars in this mass range.
If the orbital inclination
at the time of the spectroscopic observations presented here was as low as 
$40^\circ$, this would imply true stellar mass of $1.7$ M$_\odot$, 
which corresponds to an effective temperature of $\sim 4670$ K, or a
spectral type between K3 and K4, using the \citet{sie00} evolutionary
tracks at 2 Myr.  This is very inconsistent with the observed spectral type.
Thus, we infer that the orbital inclination at the time of the spectroscopic 
observations could not be much less than $61.^\circ8$.

{\bf \subsubsection{Production of Excess H$\alpha$ Emission}}

If the excess H$\alpha$ emission does come from the planetary
companion, how is it produced?  One of the most remarkable aspects of the 
apparent excess H$\alpha$ emission is its strength.  Tables 1 and 2
show that, when visible, the planet's H$\alpha$ equivalent width is typically
70\% to 80\% that from the star.  Because the excess equivalent width and the
stellar chromospheric equivalent width are measured relative to the same 
stellar continuum, the H$\alpha$ luminosity from the planet then reaches 80\%
that from the star.  The 10.5 \AA\ emission equivalent width from the star, 
while stronger than the dM3e star AD Leo as shown in Figure 3, is not 
atypical of very magnetically active dM3-4e stars, several of
which have H$\alpha$ emission equivalent widths of 10.0 \AA\ or more 
\citep[e.g.][]{haw96}.  However, the planet 
orbiting PTFO8-8695 is much smaller than
the star - the apparent area of the planet is $\sim 3.4$\% 
that of the star \citep{vey12}.  In
order to produce an H$\alpha$ equivalent width that is $\sim 75$\% that of 
the star, the H$\alpha$ surface flux of the planet would have to be 22 times
that of the star.  We do not know the effective temperature or spectral type
of the young planet orbiting PTFO8-8695; however, it is likely cooler and
later in type than the star.  Such a dramatic rise in H$\alpha$ surface flux
is not observed in active stars or brown dwarfs with spectral types later 
than M4 \citep[e.g.][]{sch10, wes11}.  As a result, it is doubtful the 
emission from the planet around PTFO8-8695 is produced by magnetic activity. 

We further argue that the emission from this planet does not arise on its
surface at all, but instead either in a confined region around the planet
such as in the planetary magnetosphere, or in a more complicated flow with
the planet feeding the material.  Similar gas flows have been observed in
absorption \citep[e.g.][]{cau15, ehr15} and have been modeled by a number of
investigators \citep[e.g.][]{mat15}.
The radius of the planet is in the range of 1.64 
R$_{JUP}$ \citep{bar13} to 1.91 R$_{JUP}$ \citep{vey12}.  Assuming the 
planet's rotation is tidally locked with its orbit (the synchronization
timescale is estimated to be $<< 1$ Myr using relationships given in Gu et 
al. 2003), the maximum $v$sin$i$ 
the planet could have is 18.6 -- 21.7 km s$^{-1}$.  However, the average 
velocity width of the excess H$\alpha$ emission is $87.3 \pm 4.9$ km s$^{-1}$,
significantly more than can be accounted for by the rotation of the planet.  
Instead, we suggest the most likely explanation is that the emission results
from mass outflow from the planet.  

{\bf \subsubsection{Mass Loss from the Companion}}

Mass outflow from a hot Jupiter was
first detected by \citet{vid03} and these observations have been confirmed
by several additional studies \citep[e.g.,][]{lin10}.  Mass loss from hot
Jupiters has been studied theoretically by a number of investigators 
\citep[e.g.,][]{lam03, bar04, bar06, mur09, lai10, tra11, ada11}.
\citet{vey12} noted that this planet's 
radius is very close to its Roche lobe radius.  Using equation (1) of 
\citet{vey12} and the best fitting parameters of \citet{bar13} actually 
places the Roche radius of the planet slightly inside the inferred radius 
of the planet.  Thus, the potential planet orbiting PTFO8-8695 is a prime
candidate for significant mass loss.  Using the best fit parameters from 
\citet{bar13}, the escape velocity from the planet is 82 -- 88 km s$^{-1}$ 
which is very comparable to the velocity width of the excess H$\alpha$ 
emission apparently associated with the planet, further suggestive that
this planet may be losing considerable mass.  

\citet{mur09} was the first to 
theoretically study mass loss from a hot Jupiter orbiting a young 
pre-main sequence star.  These authors note that the very high mass loss
rates from T Tauri stars will likely completely stifle planetary flows on the
day side of the hot Jupiter as the result of the large ram pessure associated
with the stellar wind.  However, the large mass loss rates used by
\citet{mur09} are only really appropriate for CTTSs which are still
accreting material from a disk.  Stellar wind mass loss rates appropriate for
a WTTS such as PTFO8-8695 have not yet been determined.
The semi-major axis of the suspected companion to PTFO8-8695 is less
than $\sim 2$ $R_\odot$ while the star itself has a radius of $\sim 1$
$R_\odot$ \citep{vey12, bar13}.  Thus, the planet would be orbiting only
about 1 stellar radius above the stellar surface.  T Tauri stars are
measured to have average surface magnetic fields of 2 -- 3 kG \citep{joh07}, 
so it is likely
that the planet is orbiting inside the Alfv\'en radius of the star, in
which case \citet{mur09} suggest the outflow from both the day and night
side of the planet may be suppressed.  However, the planetary magnetosphere
may help balance the stellar magnetic pressure in order to allow the
wind from the planet 
to get started, and it may also lead to the outflowing material 
building up substantial optical depth at a size that is several times
the nominal planetary radius \citep[e.g.,][]{tra11} which is needed in
order to produce the amount of H$\alpha$ emission seen as discussed above.  
Once the material escapes the planet and its magnetosphere, it may then 
be the stellar field that ultimately controls
where the material flows.

Most models of mass loss from a hot Jupiter find that the flow ultimately 
gets redirected by the stellar wind with the material eventually leaving the
system.  Here though, the planetary wind would be launched inside the region
of space governed by the stellar magnetosphere.  In the case of CTTSs, 
much of the material flowing off the disk near the co-rotation radius is
forced by the stellar magnetosphere to travel along the field lines and
accrete onto the star \citep[e.g.,][]{bou07}.  \citet{vey12} find
that the stellar rotation period is locked to the candidate planet's orbital 
period, so the planet would be feeding material into the stellar magnetosphere
at the co-rotation radius which is very analogous to CTTSs accretion models 
\citep[e.g.,][]{shu94}.  It may then be that material flowing off the 
planet is accreting onto the star in a magnetospheric accretion flow similar
to that in CTTSs.  This accreting material might then produce emission at higher
redshifted velocities as it accelerates in the gravitational potential
well of the star, as modeled for example by \citet{mat15}.  
Such higher velocity tails are hinted at in Figures
7 and 8.  Material that is lost into a wind may also contribute 
H$\alpha$ emission at velocities different from the orbital velocity of
the suspected planet.

{\bf \subsubsection{A Paradigm to Explain the Full Range of H$\alpha$ Profiles}}

The interplay of material being lost by the planet embedded in the 
magnetosphere of a young star might explain all the H$\alpha$ variations
observed in PTFO8-8695.  Line profiles seen in Figures 1, 5, and 6 suggest
that there are times when only emission from the chromospherically active
star is present.  As mentioned above, PTFO8-8695 is known to
flare.  A strong stellar flare may rapidly strip away the outflowing
material from the planet resulting in no excess emission for some time.
Due to flares or other dynamo related variability, the planet at times may 
be in regions dominated by open stellar field lines where a
strong stellar wind is flowing which stifles the planetary outflow as
suggested by \citet{mur09}.  At such times, there may be no excess 
H$\alpha$ emission associated with the planet.  At some point after a flare
or once the planet moves into a region of closed stellar fields, the 
planetary outflow may be re-established and we observe
the excess emission associated with the planet.  As the flow continues and
material begins to fill the stellar magnetosphere, accrete onto the
star, or be lost to a wind,
H$\alpha$ emission will begin to have contributions from gas no
longer directly tied to the planet.  The radial velocity structure of this
gas as viewed from Earth, and the resulting line profile, might then get
quite complex and produce the various profile shapes observed in Figure 1
and 9,
as well as producing the higher velocity tails seen in Figures 5 and 6.
The possibility that some component of the H$\alpha$ excess emission is
not directly tied to the planet cautions that the eccentric Keplerian
orbit discussed above may not be real, but simply due to the difficulty
in isolating the emission coming directly from the planet.  Indeed,
since the RV variations traced by the excess H$\alpha$ emission likely
include components not strictly in orbit with the suspected planet, it is
difficult to definitively rule out any of the specific orbital fits shown 
in Figure 10.

When the stellar field adjusts itself due to another flare or some dynamo
variation, the sequence outlined above could then start all over again.
\citet{bar13} suggest that the planet's orbital plane is inclined relative
to the stellar equatorial plane, so
even though the stellar rotation and planet's orbit may be locked, the
planet is still moving in latitude relative to the star and hence is
moving through different parts of the stellar magnetosphere.  This motion
through the stellar magnetic field could
produce changes in a planetary flow within even a single orbit, leading
to complicated variations in the H$\alpha$ line profile.  To explain the
full range of variation in the line profiles observed from PTFO8-8695,
we are forced to invoke accretion onto the star.  As discussed above,
there is no suggestion of an infrared excess in this star, even out to
24 $\mu$m.  Feeding the accretion flow with material escaping from a planet
naturally explains the lack of an infrared excess, and the profile 
variability in PTFO8-8695 is at times well matched assuming excess H$\alpha$
emission that is physically associated with the reported planet around this
star.  Thus, we find the planet scenario outlined above
the most likely explanation for the variations
we see in the lines.  Above, we crudely estimated that the excess H$\alpha$
emission would imply an accretion rate of $3 \times 10^{-10}$ M$_\odot$
yr$^{-1}$ onto the star.  If this accretion rate is fed entirely by a
planet of $\sim 7$ M$_{JUP}$ as suggested above, the lifetime of the 
planet would be $\sim 2 \times 10^{7}$ yr.  Thus, if the candidate
planet orbiting PTFO8-8695 is real, we expect it to evolve substantially
over the pre-main sequence lifetime of this star.

%We do note that the escape velocity from the star at
%the planet's orbital distance is 272 -- 311 km s$^{-1}$ using the
%preferred values in \citet{bar13}.  Such a large velocity width is not
%observed in the excess H$\alpha$ emission line, so the ultimate fate of
%material being lost from the planet is likely that is accretes onto the
%central star.

\section{Summary}

We have used relatively high time cadence, high spectral resolution
optical observations to detect excess H$\alpha$ emission from the $2 - 3$ 
Myr old WTTS PTFO8-8695.  At some times these high cadence observations
show that the excess emission appears to move in velocity as expected if 
it were produced by the suspected planetary companion to this young
star.  We have considered the possibility that the observed excess emission 
is produced by stellar activity (flares), accretion from a disk, or from a 
planetary companion; we find the planetary companion to be the most likely 
explanation.  Yu et al. (2015) recently examined additional photometry 
and spectroscopy of this star in an effort to test the planet hypothesis for 
this system.  They do not favor the hot Jupiter hypothesis, instead suggesting
that their data point to either starspots, eclipses by circumstellar dust 
(fed by either a circumstellar disk or a low mass evaporating planet), or 
occultations of an accretion hotspot as the most likely explanation for the 
variations they observe.  However, these authors did not observe the strong, 
variable, excess Balmer emission that we discuss here and their alternatives 
would not account for it.  Above, we discuss the difficulties associated with
a stellar activity or disk accretion origin for the H$\alpha$ variations, and
we conclude that an evaporating planet is the best explanation for the 
variations we observe.  While no single model may fit all the data on this 
star, this may be due to the extreme nature of this object as a very rapidly
rotating, magnetically active pre-main sequence star.  Therefore, we believe
the planetary companion hypothesis is still a viable component of this unique
system.

If the excess H$\alpha$ emission we see does come from a planetary
companion, the strength of the emission indicates that it arises in 
an extended volume around the planet, likely fed by mass loss from the
planet which is expected to be overflowing its Roche lobe.  Interpreting
the radial velocity variations of the excess H$\alpha$ emission as coming
from the planet, we place an upper limit on the mass of the star as 
$M_* {\rm sin}^3i = 0.535 \pm 0.047$ M$_\odot$, while the planet's
mass would then be $M_P {\rm sin}^3i = 4.45 \pm 0.34$ M$_{JUP}$.  While
there is evidence that the orbital inclination of this system varies
dramatically due to nodal precession \citep{bar13, cia15}, the inclination
at the time of our spectroscopic observations can not be too low or the 
stellar mass would become very inconsistent with its spectral type.  This
leads to an upper limit for the planet's mass of $\sim 7$ M$_{JUP}$.  While
the observations presented here are highly suggestive that we have directly
observed the spectroscopic signature of a mass losing planet in orbit around
this young star, these results are primarily based on relatively low 
signal-to-noise observations.  Further high signal-to-noise, high cadence
time resolved
H$\alpha$ observations over several predicted orbits will greatly aid in
solidifying the picture outlined in this work.  In addition,
we caution that theoretical work is likely still needed
to determine whether such strong emission could really be produced by a
young hot Jupiter undergoing mass loss.

\acknowledgements

LP and CMJ-K wish to acknowledge partial support for this research from 
the NASA Origins of Solar Systems program through grant number 
07-SSO07-86 made to Lowell Observatory.  CMJ-K also wishes
to acknowledge partial support for this work from the National Science
Foundation through grant number 1212122 made to Rice University.  This work 
was funded in part through the 2013-2014 NAU/NASA Space Grant Undergraduate 
Research Internship for Jacob N. McLane.  Finally, we wish to thank an
anonymous referee for several useful comments that improved the original
manuscript.

\clearpage
 
\begin{deluxetable}{lcccccc}
%\rotate
\tablewidth{15.0truecm}   %8.8truecm maximum
\tablecaption{McDonald Observing Log for 15 November 2013\label{obslog}}
\tablehead{
   \colhead{UT}&
   \colhead{}&
   \colhead{Excess H$\alpha$}&
   \colhead{Excess H$\alpha$}&
   \colhead{Excess H$\alpha$}&
   \colhead{}&
   \colhead{Pred. Planet}\\[0.2ex]
   \colhead{Time\tablenotemark{a}}&
   \colhead{S/N\tablenotemark{b}}&
   \colhead{$W_{eq}$ (\AA )}&
   \colhead{$\sigma_D$ (km s$^{-1}$) }&
   \colhead{RV (km s$^{-1}$) }&
   \colhead{Phase\tablenotemark{c}}&
   \colhead{RV (km s$^{-1}$)\tablenotemark{c} }
}
\startdata
05:43 & 3.2 & $8.8 \pm 0.3$ & 86.5 & $189.6 \pm 9.4$ & 0.282 & 195.1 \\
08:07 & 4.3 & $9.2 \pm 0.2$ & 120.3 & $34.4 \pm 4.0$ & 0.505 & -6.2 \\
08:53 & 4.7 & $7.3 \pm 0.2$ & 104.4 & $-58.8 \pm 4.2$ & 0.577 & -98.2 \\
09:44 & 5.7 & $5.6 \pm 0.2$ & 59.3 & $-156.5 \pm 6.1$ & 0.656 & -164.8 \\
10:32 & 5.1 & $8.2 \pm 0.2$ & 84.8 & $-175.4 \pm 5.9$ & 0.730 & -197.1 \\
11:17 & 4.4 & $8.2 \pm 0.2$ & 80.9 & $-190.6 \pm 7.7$ & 0.800 & -189.0 \\
\enddata
\tablenotetext{a}{UT time at the midpoint of the exposure.}
\tablenotetext{b}{Signal-to-noise per pixel in the continuum near H$\alpha$.}
\tablenotetext{c}{Based on ephemeris in \citet{vey12}.}
\end{deluxetable}

\clearpage
 
\begin{deluxetable}{lcccccccc}
%\rotate
\tablewidth{18.0truecm}   %8.8truecm maximum
\tablecaption{Kitt Peak Observing Log\label{obslog2}}
\tablehead{
   \colhead{UT}&
   \colhead{}&
   \colhead{Ind. Exp.}&
   \colhead{}&
   \colhead{Excess H$\alpha$}&
   \colhead{Excess H$\alpha$}&
   \colhead{Excess H$\alpha$}&
   \colhead{}&
   \colhead{Pred. Planet}\\[0.2ex]
   \colhead{Time\tablenotemark{a}}&
   \colhead{$N_{exp}$}&
   \colhead{Time (s)}&
   \colhead{S/N\tablenotemark{b}}&
   \colhead{$W_{eq}$ (\AA )}&
   \colhead{$\sigma_D$ (km s$^{-1}$) }&
   \colhead{RV (km s$^{-1}$) }&
   \colhead{Phase\tablenotemark{c}}&
   \colhead{RV (km s$^{-1}$)\tablenotemark{c} }
}
\startdata
      &   &     &     &               & 8 Dec 2012 &                 &       &       \\
04:33 & 3 & 600 & 7.2 & \nodata & \nodata & \nodata  & 0.484 & 20.0 \\
05:16 & 3 & 600 & 6.9 & \nodata & \nodata & \nodata  & 0.551 & -62.7 \\
06:02 & 3 & 600 & 6.8 & \nodata & \nodata & \nodata  & 0.621 & -137.2 \\
06:46 & 3 & 600 & 7.7 & \nodata & \nodata & \nodata  & 0.689 & -184.7 \\
07:29 & 3 & 600 & 7.8 & \nodata & \nodata & \nodata  & 0.756 & -199.0 \\
08:13 & 3 & 600 & 5.8 & \nodata & \nodata & \nodata  & 0.824 & -178.0 \\
09:14 & 3 & 600 & 8.7 & \nodata & \nodata & \nodata  & 0.919 & -97.0 \\
09:57 & 3 & 600 & 7.4 & \nodata & \nodata & \nodata  & 0.986 & -17.5 \\
10:39 & 3 & 600 & 7.9 & \nodata & \nodata & \nodata  & 0.051 & 62.7 \\
11:21 & 3 & 600 & 7.2 & \nodata & \nodata & \nodata  & 0.115 & 131.7 \\
11:55 & 2 & 600 & 6.1 & \nodata & \nodata & \nodata  & 0.169 & 173.9 \\
      &   &     &     &               &      &                 &       &       \\
      &   &     &     &               & 9 Dec 2012 &                 &       &       \\
04:35 & 3 & 600 & 5.5 & \nodata & \nodata & \nodata & 0.719 & -195.3 \\
05:23 & 3 & 600 & 7.4 & \nodata & \nodata & \nodata & 0.792 & -192.2 \\
06:11 & 3 & 600 & 6.2 & \nodata & \nodata & \nodata & 0.867 & -147.7 \\
06:55 & 3 & 600 & 2.6 & \nodata & \nodata & \nodata & 0.935 & -79.1 \\
07:42 & 3 & 600 & 3.1 & \nodata & \nodata & \nodata & 0.008 & 10.0 \\
08:21 & 2 & 600 & 6.4 & \nodata & \nodata & \nodata & 0.068 & 82.5 \\
09:00 & 3 & 600 & 4.1 & \nodata & \nodata & \nodata & 0.128 & 143.4 \\
09:47 & 3 & 600 & 6.8 & \nodata & \nodata & \nodata & 0.201 & 189.7 \\
10:38 & 3 & 600 & 5.5 & \nodata & \nodata & \nodata & 0.280 & 195.6 \\
11:27 & 3 & 600 & 6.8 & \nodata & \nodata & \nodata & 0.356 & 156.6 \\
      &   &     &     &               &      &                 &       &       \\
      &   &     &     &               & 10 Dec 2012 &                 &       &       \\
04:25 & 1 & 1200 & 4.4 & $0.3 \pm 0.2$ & \nodata & \nodata & 0.902 & -115.0 \\
04:48 & 1 & 1200 & 5.2 & $-0.1 \pm 0.2$ & \nodata & \nodata & 0.931 & -83.6 \\
05:27 & 1 & 1200 & 7.1 & $1.8 \pm 0.2$ & 83.4 & $57.3 \pm 50.0$ & 0.968 & -39.8 \\
05:53 & 1 & 1200 & 7.3 & $3.6 \pm 0.2$ & 92.3 & $99.4 \pm 15.2$ & 0.026 & 32.4 \\
06:15 & 1 & 1200 & 8.5 & $4.7 \pm 0.2$ & 111.4 & $174.0 \pm 11.5$ & 0.068 & 82.5 \\
06:45 & 1 & 1200 & 7.2 & $6.4 \pm 0.2$ & 101.6 & $272.4 \pm 10.9$ & 0.102 & 160.3 \\
07:12 & 1 & 1200 & 8.1 & $5.7 \pm 0.2$ & 97.2 & $265.9 \pm 10.0$ & 0.149 & 185.1 \\
07:34 & 1 & 1200 & 7.0 & $4.7 \pm 0.2$ & 88.6 & $251.7 \pm 12.6$ & 0.225 & 196.6 \\
08:03 & 1 & 1200 & 6.0 & $3.8 \pm 0.2$ & 86.4 & $245.9 \pm 17.2$ & 0.269 & 197.7 \\
08:25 & 1 & 1200 & 7.1 & $2.7 \pm 0.2$ & 62.3 & $202.4 \pm 16.8$ & 0.303 & 188.2 \\
08:47 & 1 & 1200 & 7.6 & $2.6 \pm 0.2$ & 69.1 & $198.8 \pm 16.3$ & 0.337 & 170.1 \\
\enddata
\tablenotetext{a}{UT time at the midpoint of exposure(s).}
\tablenotetext{b}{Signal-to-noise per pixel in the continuum near H$\alpha$.}
\tablenotetext{c}{Based on ephemeris in \citet{vey12}.}
\end{deluxetable}

\clearpage

%figure 1
\begin{figure} 
\epsscale{.80}
\plotone{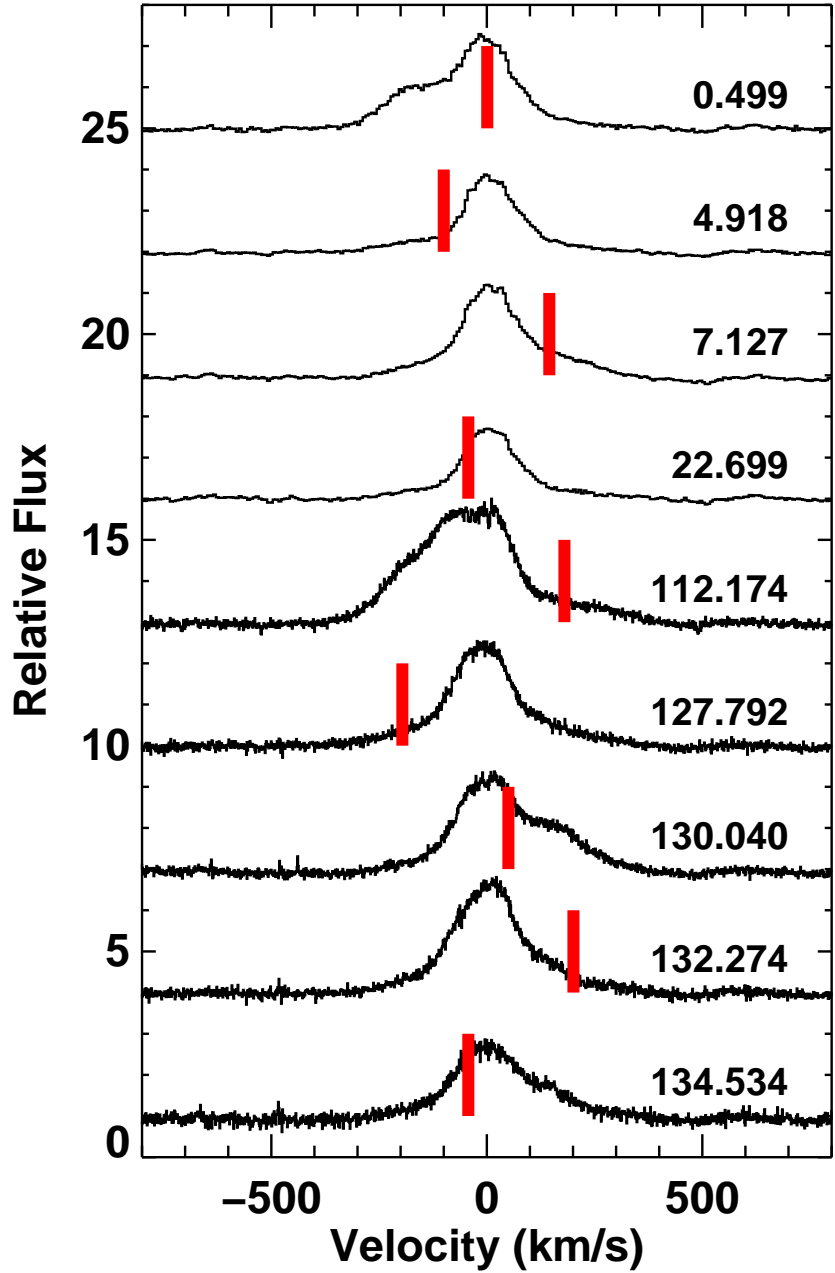}
\caption{The continuum normalized H$\alpha$ profiles of PTFO8-8695 obtained 
with the HET (top 4 profiles) and with Keck (bottom 5 profiles), as part
of the discovery study of \citet{vey12}.  The orbital phase (with 0.0 the
midpoint of the first transit to precede the first HET observation) at the 
midpoint of the observation is given with each profile.  
The thick, red vertical line
on each profile marks the expected velocity position of the planetary
companion detected by \citet{vey12} from transit observations of this
WTTS.  There is clear variability in the H$\alpha$ emission.}
\end{figure}

\clearpage

%figure 2
\begin{figure} 
\epsscale{.90}
\plotone{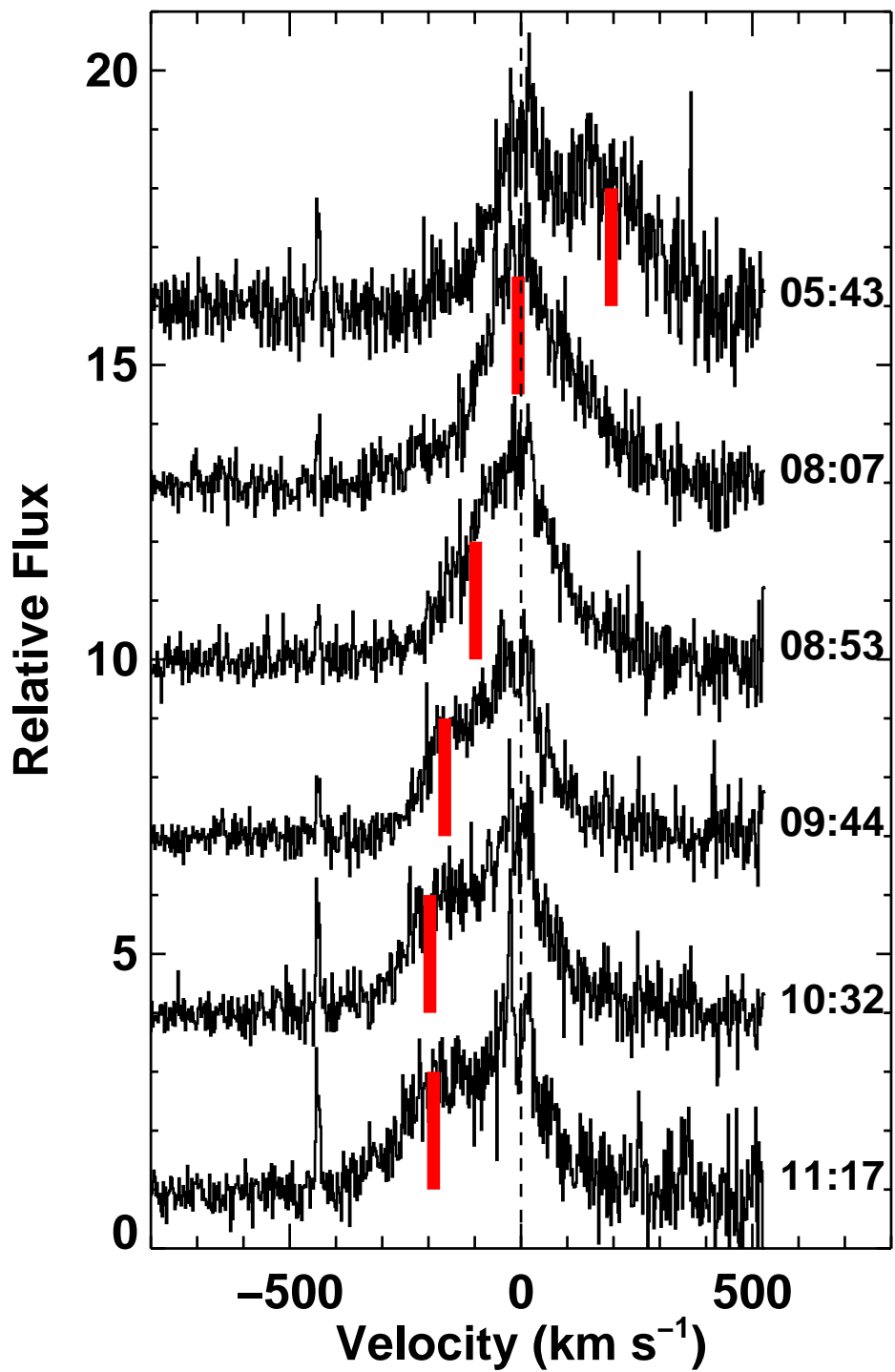}
\caption{The 6 observed continuum normalized H$\alpha$ profiles of 
PTFO8-8695 obtained on 15 November 2013 UT at McDonald Observatory
are shown with the UT time of the midpoint of each observation given.  
The thick, red vertical line
on each profile marks the expected velocity position of the planetary
companion detected by \citet{vey12} from transit observations of this
WTTS.  There is clear excess H$\alpha$ emission coincident
in velocity space at the radial velocity of the planet. }
\end{figure}

\clearpage

%figure 3
\begin{figure} 
\epsscale{.90}
\plotone{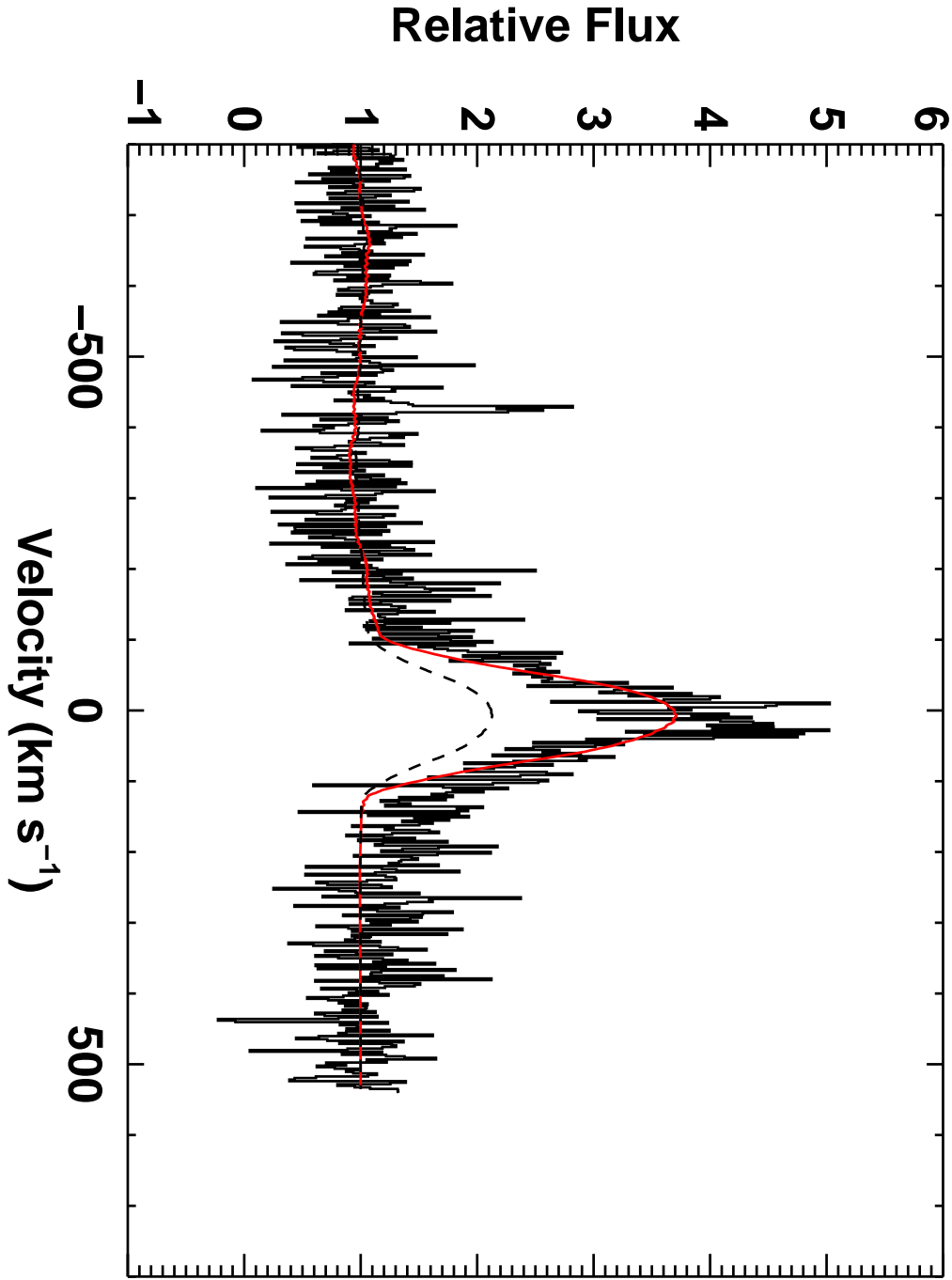}
\caption{The black histogram shows our estimate of the stellar (chromospheric)
H$\alpha$ emission produced by the WTTS PTFO8-8695 in the
continuum normalized spectrum.  The dashed line shows a spectrum of AD Leo
(dM3e) rotationally broadened to the same $v$sin$i$ as PTFO8-8695.  The smooth
red profile shows the spectrum of AD Leo in which the H$\alpha$ emission has
been multiplied by a factor 2.4 after first rotationally broadening the
spectrum to the same $v$sin$i$ as PTFO8-8695.}
\end{figure}

\clearpage

%figure 4
\begin{figure} 
\epsscale{.90}
\plotone{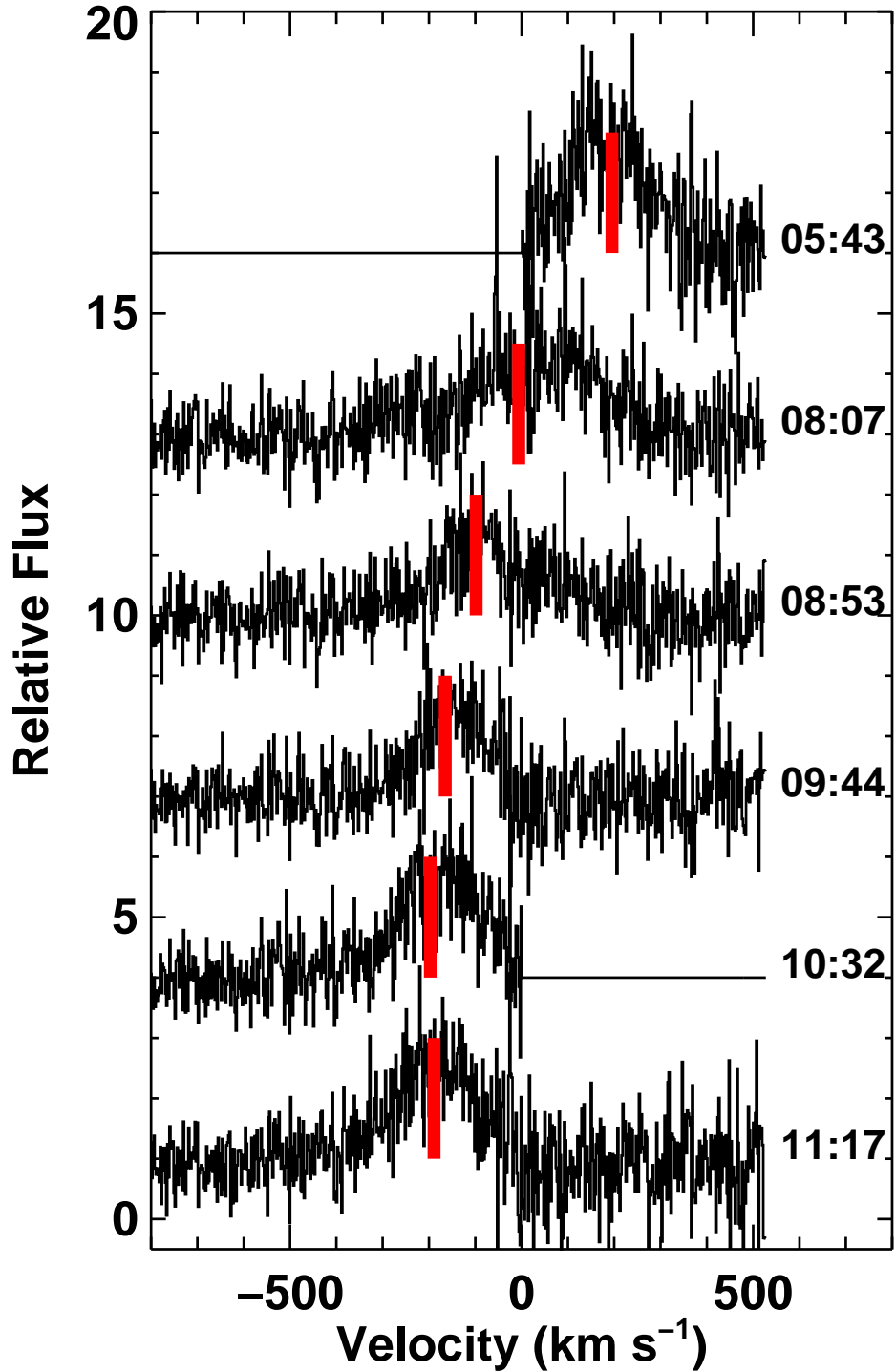}
\caption{The 6 observed continuum normalized H$\alpha$ profiles of 
PTFO8-8695 obtained on 15 November 2013 UT with the stellar (chromospheric)
component subtracted out are plotted, again with the UT time of the midpoint
of each observation given.  
The thick, red vertical line on each profile again
marks the expected velocity position of the planetary
companion detected by \citet{vey12} from transit observations of this
WTTS.}
\end{figure}

\clearpage

%figure 5
\begin{figure} 
\epsscale{.80}
\plotone{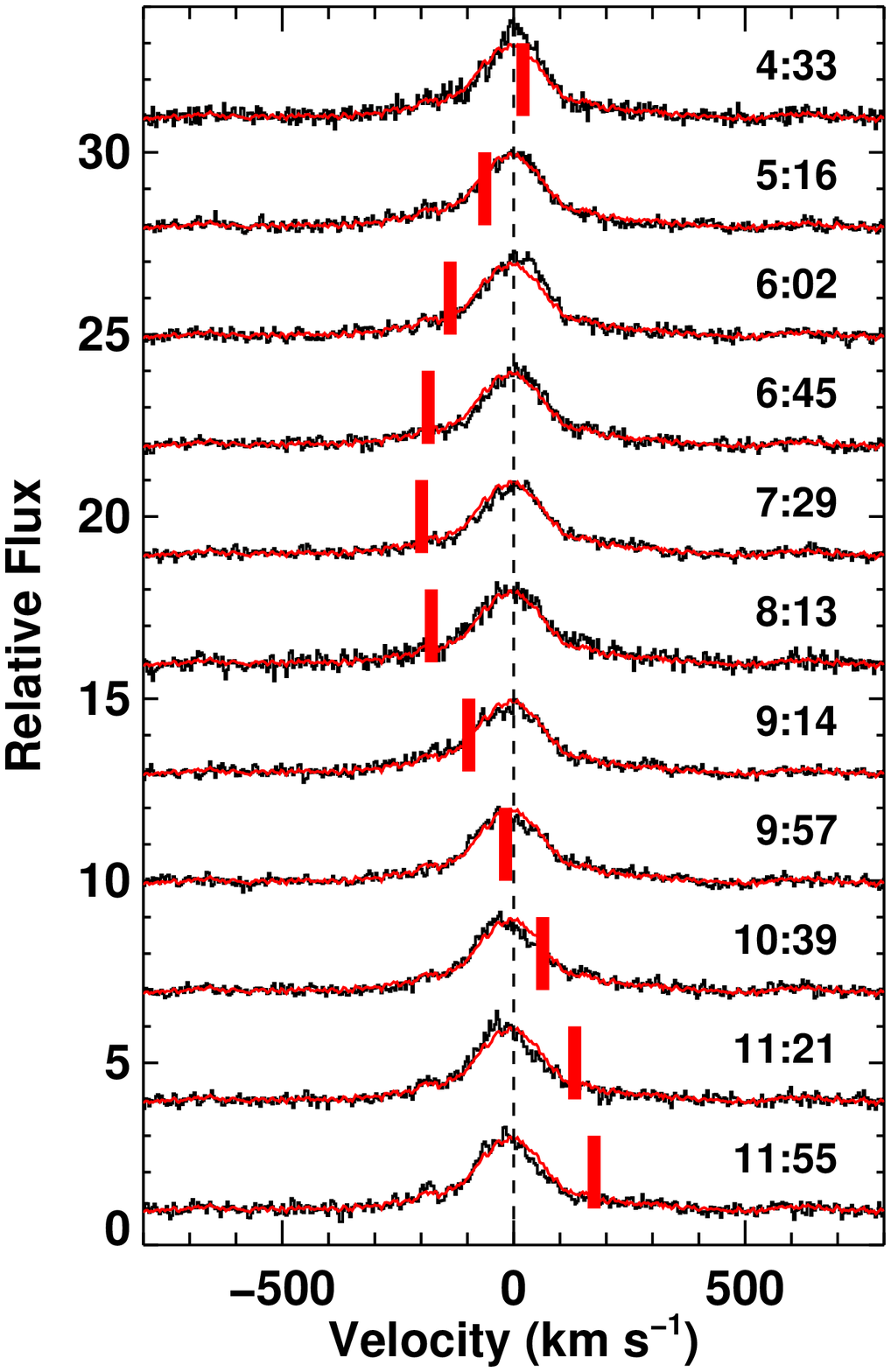}
\caption{The 11 observed continuum normalized H$\alpha$ profiles of 
PTFO8-8695 obtained on 8 December 2012 UT at Kitt Peak Observatory
are shown in the black histograms with the UT time (midpoint) of each observation.
The smooth red curve shows the
average H$\alpha$ profile from this night.  The thick, red vertical line
on each profile marks the expected velocity position of the planetary
companion detected by \citet{vey12} from transit observations of this
WTTS. No clear excess emission is seen in the line profile on this night.}
\end{figure}

\clearpage

%figure 6
\begin{figure} 
\epsscale{.80}
\plotone{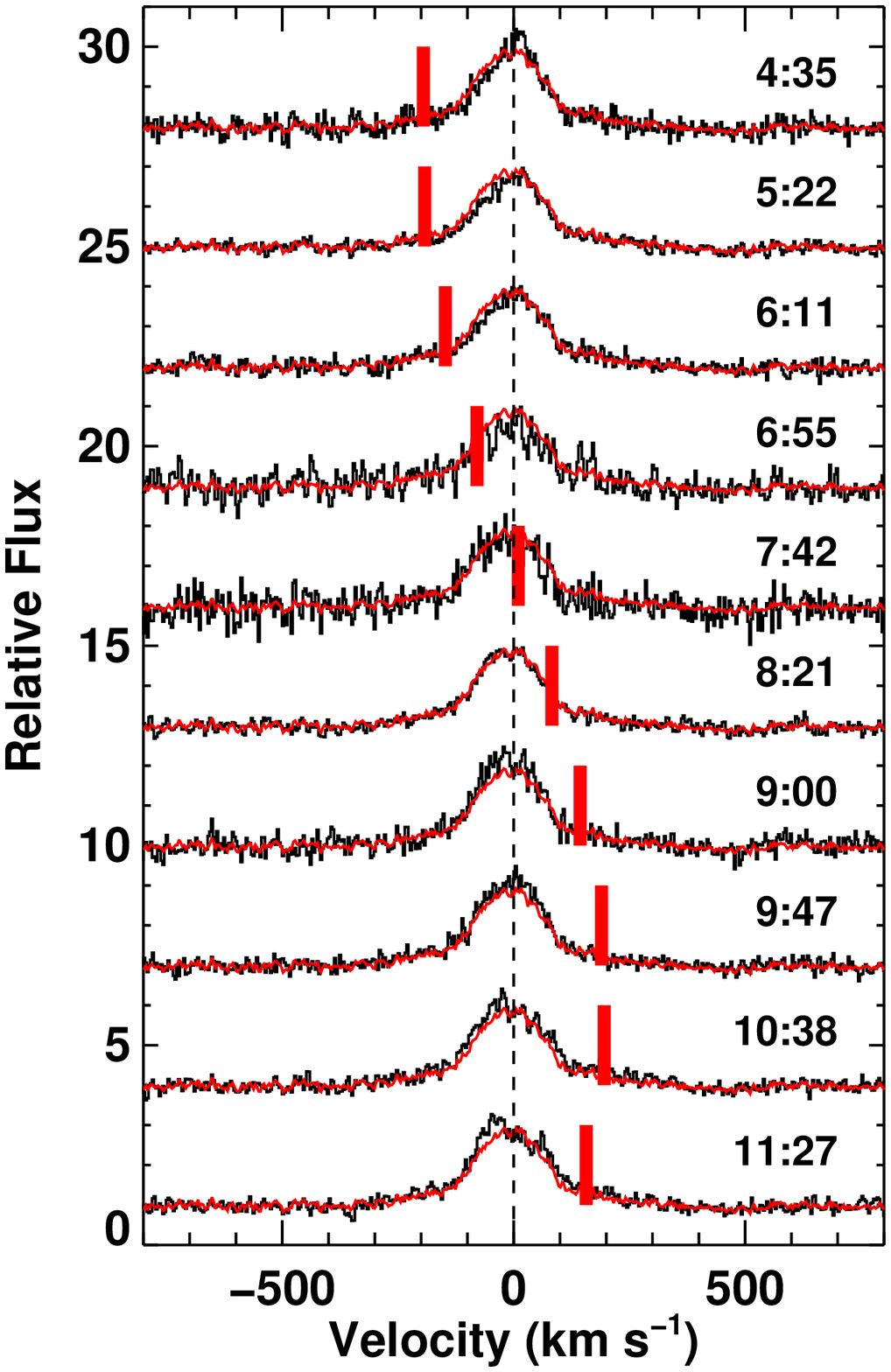}
\caption{The 10 observed continuum normalized H$\alpha$ profiles of 
PTFO8-8695 obtained on 9 December 2012 UT at Kitt Peak Observatory
are shown in the black histograms with the UT (midpoint) time of each observation.
The smooth red curve shows the
average H$\alpha$ profile from this night.  The thick, red vertical line
on each profile marks the expected velocity position of the planetary
companion detected by \citet{vey12} from transit observations of this
WTTS. No clear excess emission is seen in the line profile on this night.}
\end{figure}

\clearpage

%figure 7
\begin{figure} 
\epsscale{.80}
\plotone{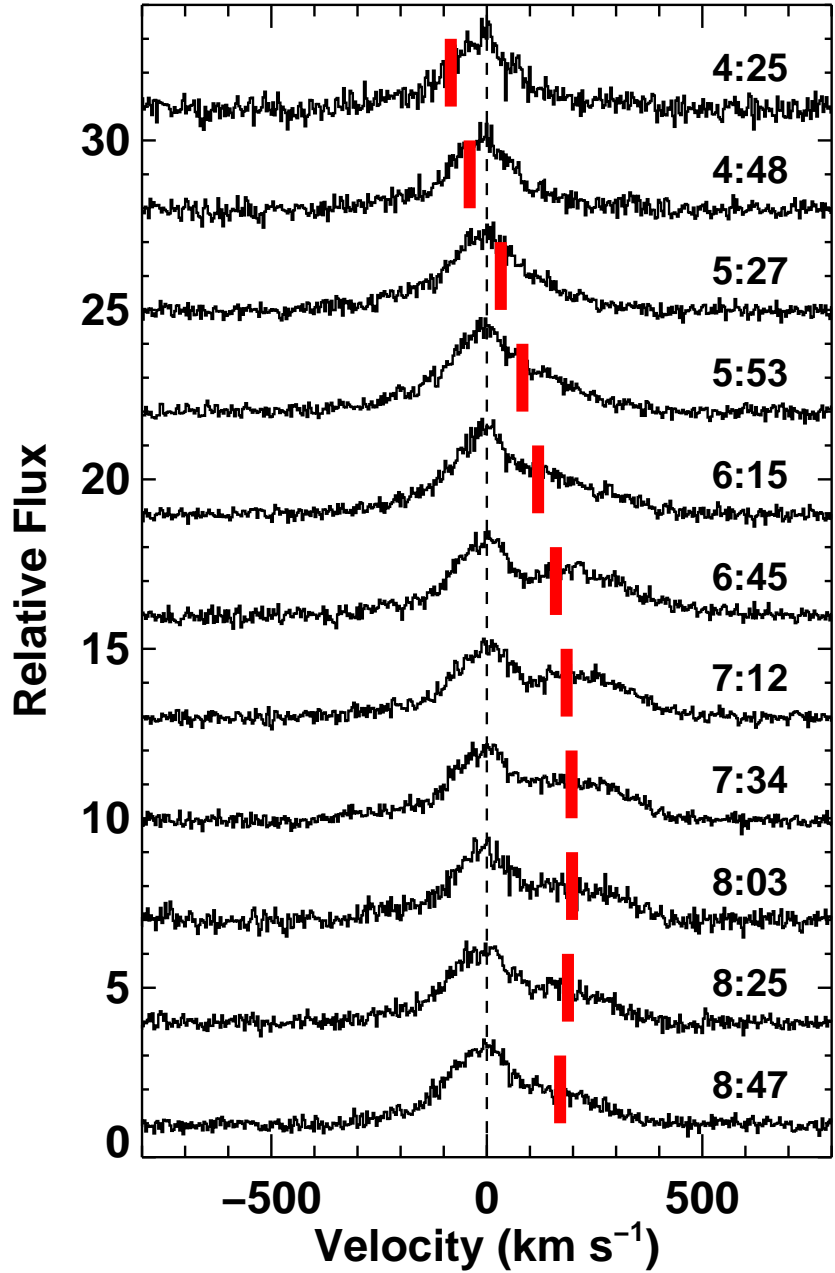}
\caption{The 11 observed continuum normalized H$\alpha$ profiles of 
PTFO8-8695 obtained on 10 December 2012 UT at Kitt Peak Observatory are
plotted with the UT time (midpoint) of each observation.  The thick, red vertical line
on each profile marks the expected velocity position of the planetary
companion detected by \citet{vey12} from transit observations of this
WTTS.  Starting at UT 5:53 there is clear excess H$\alpha$ emission coincident
in velocity space at the radial velocity of the planet. }
\end{figure}

\clearpage

%figure 8
\begin{figure} 
\epsscale{.80}
\plotone{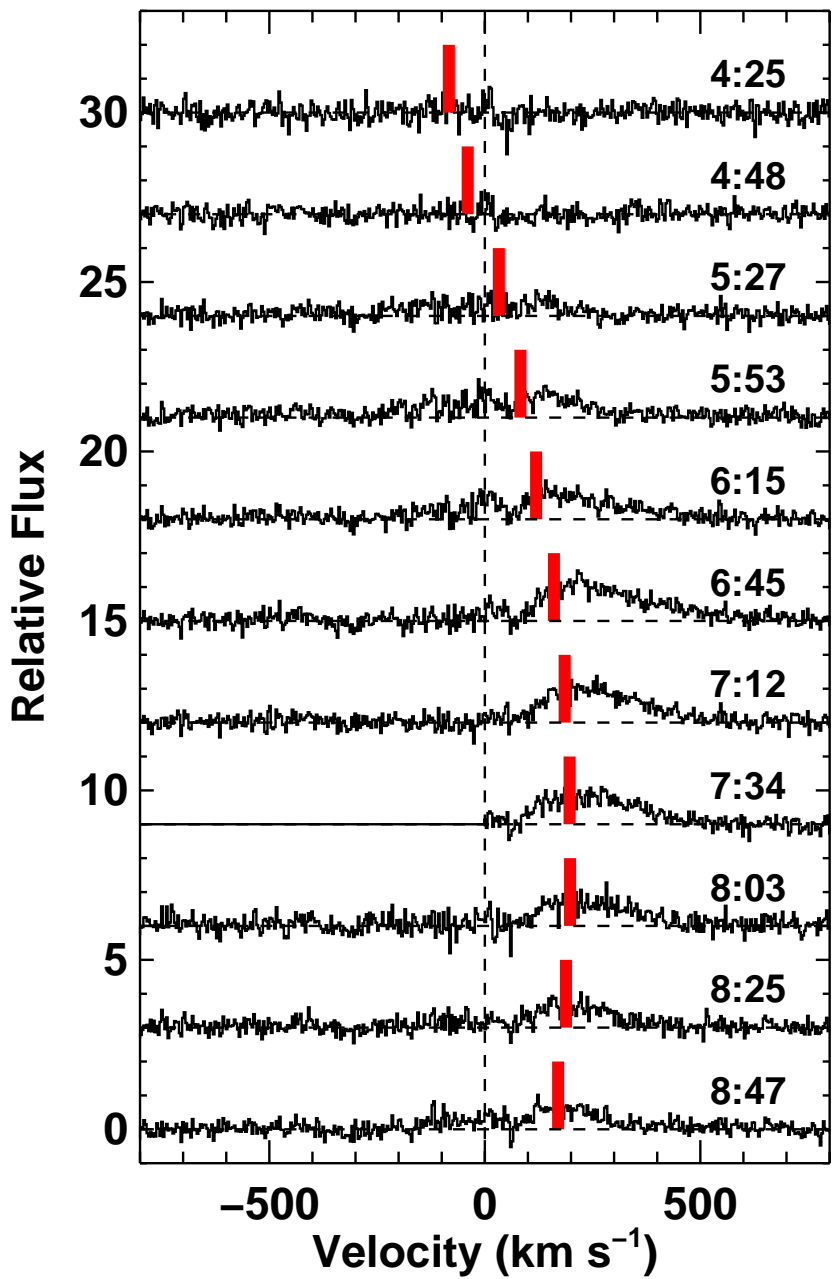}
\caption{The 11 observed continuum normalized H$\alpha$ profiles of 
PTFO8-8695 obtained on 10 December 2012 UT with the stellar (chromospheric) 
component subracted out are plotted.
The thick, red vertical line
on each profile marks the expected velocity position of the planetary
companion detected by \citet{vey12} from transit observations of this
WTTS.  Starting at UT 5:53 there is clear excess H$\alpha$ emission coincident
in velocity space at the radial velocity of the planet. Some weak emission
also appears to be present at UT 5:27.}
\end{figure}

\clearpage

%figure 9
\begin{figure} 
\epsscale{.80}
\plotone{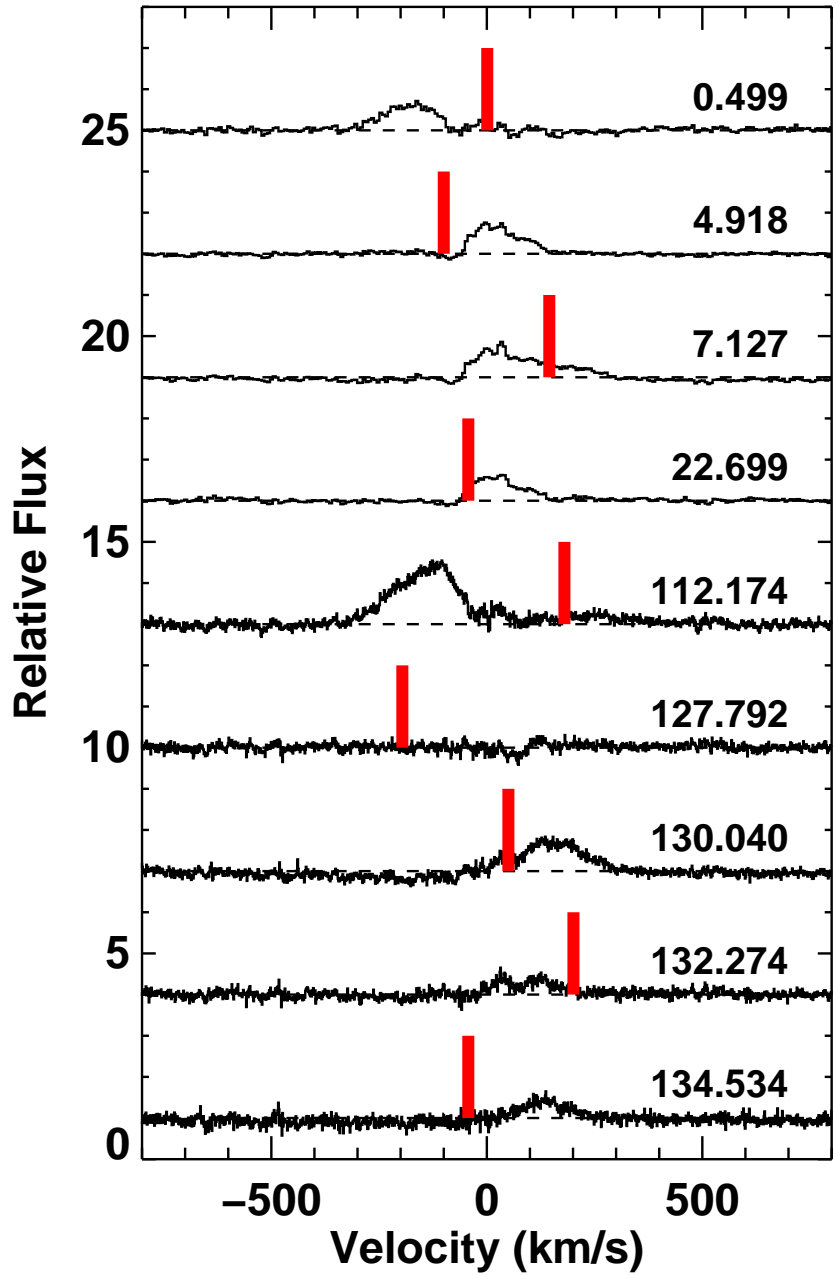}
\caption{The continuum normalized H$\alpha$ profiles of PTFO8-8695 obtained 
with the HET (top 4 profiles) and with Keck (bottom 5 profiles) from Figure 1
after an estimate of the stellar component has been subtracted off.  
The thick, red vertical line
on each profile marks the expected velocity position of the planetary
companion detected by \citet{vey12} from transit observations of this
WTTS.}
\end{figure}

\clearpage

%figure 10 
\begin{figure} 
\epsscale{.70}
\plotone{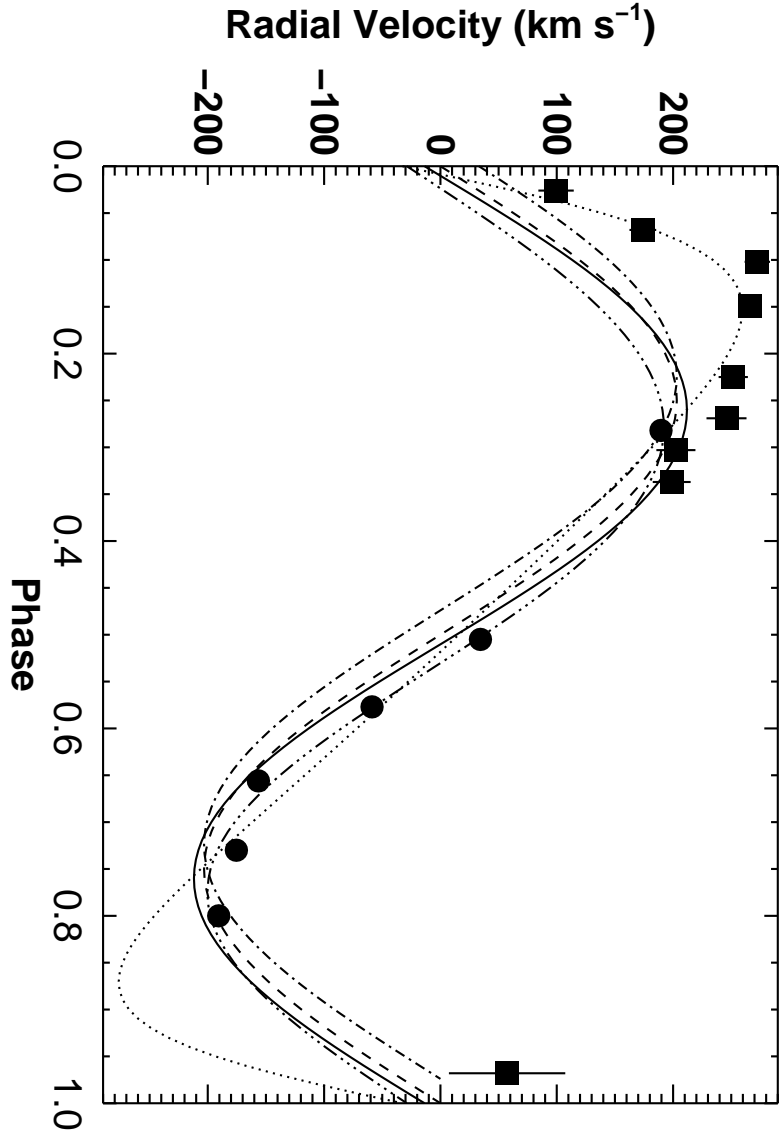}
\caption{Solid symbols show the measured RV of the excess H$\alpha$ emission
from PTFO8-8695, including uncertainties, as a function of orbital phase using
the ephemeris from \citet{vey12}.  Circles show data from McDonald Observatory
and squares show data from Kitt Peak National Observatory.  The dashed line 
shows the predicted planetary velocity curve using the ephemeris from 
\citet{vey12}.  The dash-dot line shows the predicted RV curve
from \citet{vey12} shifted in phase to account for the revised transit center
epoch determined from {\it Spitzer} data by \citet{cia15}.
The solid line is a circular orbit fit to {\bf all}
the measured RV points.  The dotted line is a full Keplerian fit to all
the measured RV points, including the transit midpoint time as a constraint 
(see text). Finally, the dash-triple dot
line is a full Keplerian fit to only
the McDonald RV points, still using the transit midpoint time as a constraint.}
\end{figure}

\end{document}